\documentclass[twocolumn, twocolappendix]{aastex631}

 \hypersetup{linkcolor=red,citecolor=blue,filecolor=cyan,urlcolor=blue}

\submitjournal{ApJS}

\shorttitle{Chemical Differentiation around Five Massive Protostars}
\shortauthors{Taniguchi et al.}

\begin{document}

\title{Chemical Differentiation around Five Massive Protostars Revealed by ALMA \\--Carbon-Chain Species, Oxygen-/Nitrogen-Bearing Complex Organic Molecules--}

\author[0000-0003-4402-6475]{Kotomi Taniguchi}
\affiliation{National Astronomical Observatory of Japan, National Institutes of Natural Sciences, 2-21-1 Osawa, Mitaka, Tokyo 181-8588, Japan}

\author[0000-0001-7031-8039]{Liton Majumdar}
\affiliation{School of Earth and Planetary Sciences, National Institute of Science Education and Research, Jatni 752050, Odisha, India}
\affiliation{Homi Bhabha National Institute, Training School Complex, Anushaktinagar, Mumbai 400094, India}

\author[0000-0003-1481-7911]{Paola Caselli}
\affiliation{Max-Planck-Institut f\"{u}r Extraterrestrische Physik, Gie{\ss}enbachstrasse 1, D-85741 Garching bei M\"{u}nchen, Germany}

\author[0000-0003-0845-128X]{Shigehisa Takakuwa}
\affiliation{Department of Physics and Astronomy, Graduate School of Science and Engineering, Kagoshima University, 1-21-35 Korimoto, Kagoshima, Kagoshima 890-0065, Japan}
\affiliation{Academia Sinica Institute of Astronomy and Astrophysics, 11F of Astro-Math Bldg., 1, Section 4, Roosevelt Road, Taipei 10617, Taiwan}

\author[0000-0002-5507-5697]{Tien-Hao Hsieh}
\affiliation{Max-Planck-Institut f\"{u}r Extraterrestrische Physik, Gie{\ss}enbachstrasse 1, D-85741 Garching bei M\"{u}nchen, Germany}

\author[0000-0003-0769-8627]{Masao Saito}
\affiliation{National Astronomical Observatory of Japan, National Institutes of Natural Sciences, 2-21-1 Osawa, Mitaka, Tokyo 181-8588, Japan}
\affiliation{Department of Astronomical Science, School of Physical Science, SOKENDAI (The Graduate University for Advanced Studies), Osawa, Mitaka, Tokyo 181-8588, Japan}

\author[0000-0002-7402-6487]{Zhi-Yun Li}
\affiliation{Department of Astronomy, University of Virginia, Charlottesville, VA 22904, USA}

\author[0000-0001-8058-8577]{Kazuhito Dobashi}
\affiliation{Department of Astronomy and Earth Sciences, Tokyo Gakugei University, Nukuikitamachi, Koganei, Tokyo 184-8501, Japan}

\author[0000-0002-1054-3004]{Tomomi Shimoikura}
\affiliation{Faculty of Social Information Studies, Otsuma Women's University, Sanban-cho, Chiyoda, Tokyo 102-8357, Japan}

\author[0000-0001-5431-2294]{Fumitaka Nakamura}
\affiliation{National Astronomical Observatory of Japan, National Institutes of Natural Sciences, 2-21-1 Osawa, Mitaka, Tokyo 181-8588, Japan}
\affiliation{Department of Astronomical Science, School of Physical Science, SOKENDAI (The Graduate University for Advanced Studies), Osawa, Mitaka, Tokyo 181-8588, Japan}
\affiliation{Department of Astronomy, Graduate School of Science, The University of Tokyo, Hongo, Bunkyo, Tokyo 113-0033, Japan}

\author[0000-0002-3389-9142]{Jonathan C. Tan}
\affiliation{Department of Astronomy, University of Virginia, Charlottesville, VA 22904, USA}
\affiliation{Department of Space, Earth \& Environment, Chalmers University of Technology, 412 93  Gothenburg, Sweden}

\author[0000-0002-4649-2536]{Eric Herbst}
\affiliation{Department of Astronomy, University of Virginia, Charlottesville, VA 22904, USA}
\affiliation{Department of Chemistry, University of Virginia, Charlottesville, VA 22904, USA}



\begin{abstract}
We present Atacama Large Millimeter/submillimeter Array Band 3 data toward five massive young stellar objects (MYSOs), and investigate relationships between unsaturated carbon-chain species and saturated complex organic molecules (COMs).
An HC$_{5}$N ($J=35-34$) line has been detected from three MYSOs, where nitrogen(N)-bearing COMs (CH$_{2}$CHCN and CH$_{3}$CH$_{2}$CN) have been detected.
The HC$_{5}$N spatial distributions show compact features and match with a methanol (CH$_{3}$OH) line with an upper-state energy around 300 K, which should trace hot cores.
The hot regions are more extended around the MYSOs where N-bearing COMs and HC$_{5}$N have been detected compared to two MYSOs without these molecular lines, while there are no clear differences in the bolometric luminosity and temperature.
We run chemical simulations of hot-core models with a warm-up stage, and compare with the observational results.
The observed abundances of HC$_{5}$N and COMs show good agreements with the model at the hot-core stage with temperatures above 160 K.
These results indicate that carbon-chain chemistry around the MYSOs cannot be reproduced by warm carbon-chain chemistry, and a new type of carbon-chain chemistry occurs in hot regions around MYSOs.
\end{abstract}

%
\section{Introduction} \label{sec:intro}

Astrochemical studies have revealed that chemical composition reflects evolutionary stages, physical conditions, and star formation processes \citep[see e.g.,][for a review]{2012A&ARv..20...56C, 2020ARA&A..58..727J}.
The most classical scenario is that unsaturated carbon-chain species (e.g., C$_{n}$H, cyanopolyynes (HC$_{2n+1}$N), C$_{n}$S) are efficiently formed mainly by gas-phase ion-molecule reactions in early starless core stage \citep[e.g.,][]{1989ApJS...69..271H, 1992ApJ...392..551S}.
These carbon-chain species were considered to be deficient in more evolved protostellar stages, because they are destroyed by reactions with ions and atomic oxygen, or adsorbed onto dust grains in dense cold prestellar cores. 
Thus, the carbon-chain species have been considered to be good chemical evolutionary indicators for early stages of star formation processes, and known as ``early-type species'' \citep{1992ApJ...392..551S,1998ApJ...506..743B}.

In contrast to the above classical scenario of carbon-chain species, a few carbon-chain-rich low-mass protostars have been discovered \citep{2008ApJ...672..371S}, and the chemical diversity was found around low-mass young stellar objects (YSOs).
The carbon-chain molecules around low-mass YSOs are formed by the initial reaction between CH$_{4}$, which sublimates from dust grains in lukewarm envelopes ($\approx25-30$ K), and C$^{+}$ \citep{2008ApJ...681.1385H}. 
Such a carbon-chain formation mechanism was named Warm Carbon-Chain Chemistry \citep[WCCC;][]{2008ApJ...672..371S}.

After the discovery of WCCC, several survey observations focusing on both carbon-chain species and saturated complex organic molecules (COMs)\footnote{COMs or iCOMs are defined as molecules consisting of more than six atoms \citep{2009ARA&A..47..427H}.} were conducted toward low-mass star-forming regions, using single-dish telescopes.
\citet{2016ApJ...819..140G} conducted survey observations of CH$_{3}$OH and C$_{4}$H with the IRAM 30m telescope.
These observations revealed that these two species are positively correlated, indicating that carbon-chain species and COMs can coexist around low-mass YSOs.
Furthermore, they found a tentative correlation between the gas-phase C$_{4}$H/CH$_{3}$OH abundance ratio and the CH$_{4}$/CH$_{3}$OH abundance ratio in ice, which supports the WCCC mechanism starting with sublimation of CH$_{4}$ \citep{2008ApJ...681.1385H}.
\citet{2018ApJS..236...52H} found that the CCH/CH$_{3}$OH column density ratios differ by two orders of magnitudes among the observed protostellar cores, and suggested that YSOs with higher CCH/CH$_{3}$OH column density ratios are possibly located near cloud edges or in isolated clouds.
These survey observations indicate the chemical diversity around low-mass YSOs.
\citet{2017ApJ...835....3L} conducted observations of H$_{2}$CO and {\it {cyclic}}-C$_{3}$H$_{2}$ ($c$-C$_{3}$H$_{2}$) toward the Ophiuchus star-forming region using the APEX 12m telescope, and suggested that $c$-C$_{3}$H$_{2}$ exists in more shielded parts of envelopes, while the H$_{2}$CO line traces mainly outer irradiated envelopes.
On the other hand, high-angular resolution observations toward two low-mass YSOs in the Ophiuchus region with the Atacama Large Millimeter/submillimeter Array (ALMA) revealed that $c$-C$_{3}$H$_{2}$ is enhanced in regions irradiated by the UV radiation from a nearby Herbig Be star, while the H$_{2}$CO and CH$_{3}$OH emission comes from shielded regions \citep{2021ApJ...922..152T}.
Thus, analyses of spatially-resolved data are necessary to investigate relationships between carbon-chain species and COMs around YSOs.
These studies about relations between carbon-chain species and COMs were progressed in nearby low-mass star-forming regions, whereas the chemical diversity in high-mass star-forming regions is still unclear.

Thanks to development of the radio interferometers including ALMA, astrochemical studies in high-mass star-forming regions have been dramatically proceeding.
Most astrochemical studies toward massive young stellar objects (MYSOs) have focused on COMs, namely hot core chemistry \citep[e.g.,][]{2019A&A...632A..57C, 2020ApJ...898...54T, 2021ApJ...907..108G, 2022MNRAS.511.3463Q}.
Regarding observations of carbon-chain species toward high-mass star-forming regions, on the other hand, single-dish surveys were performed.
For example, \citet{2020MNRAS.496.1990R} conducted survey observations toward 28 high-mass star-forming cores to investigate deuterium fractionation of HC$_{3}$N, but they could not find any significant trends with the evolutionary stages or the kinetic temperatures, which are possibly caused by low angular resolutions.
\citet{2018ApJ...854..133T, 2019ApJ...872..154T} conducted survey observations of carbon-chain species and N$_{2}$H$^{+}$ toward high-mass starless cores and high-mass protostellar objects (HMPOs).
The N$_{2}$H$^{+}$/HC$_{3}$N abundance ratio is found to decrease as cores evolve, and suggested to be an evolutionary indicator for the early stages of high-mass star-forming regions \citep{2019ApJ...872..154T}.
This downward trend of the N$_{2}$H$^{+}$/HC$_{3}$N ratio is explained by destruction of N$_{2}$H$^{+}$ by reactions with CO, and formation of HC$_{3}$N from CH$_{4}$ and/or C$_{2}$H$_{2}$.
These key species (CO, CH$_{4}$, C$_{2}$H$_{2}$) can sublimate from dust grains around 20 K, 25 K, and 50 K, respectively.
These results indicate that HC$_{3}$N forms efficiently and ubiquitously around HMPOs, or young high-mass stars embedded in dense gas and dust.
In addition, the detection rate of HC$_{5}$N around HMPOs is found to be 50\%, which is almost consistent with previous surveys toward hot cores associated with 6.7 GHz CH$_{3}$OH masers \citep{2014MNRAS.443.2252G}. 
These results suggest the presence of two chemically different envelopes around MYSOs; hot-core type and carbon-chain-rich type, which may be an analogue of the chemical diversity found around low-mass YSOs (hot corino vs. WCCC).
However, the detected HC$_{5}$N lines with low upper-state energies are likely to come from outer cold envelopes, not from inner warm envelopes or hot core regions, and then, their observational results are not enough to confirm that carbon-chain species exist around MYSOs.

A few observations focus on relationships between carbon-chain species and COMs around MYSOs by single-dish observations \citep{2017ApJ...844...68T, 2018ApJ...866..150T}. 
However, these observations were conducted using single-dish telescopes, and they could not reveal the spatial distributions of these carbon-chain species around MYSOs.
Observations using the Very Large Array (VLA) revealed the spatial distributions of cyanopolyynes (HC$_{3}$N, HC$_{5}$N, and HC$_{7}$N) in the Ka-band (36 GHz) toward the MYSO G28.28-0.36, which is known as a carbon-chain-rich/COM-poor MYSO \citep{2018ApJ...866..150T}. They found that cyanopolyynes are associated with 450\,\micron\, continuum clumps that likely contain deeply embedded low- or intermediate-mass YSOs \citep{2018ApJ...866...32T}.
Thus, they could not investigate the spatial distributions just around MYSOs using lines of cyanopolyynes in the Ka-band.
We need to observe higher upper-state-energy lines which can be excited in hot gas ($\sim 100$ K) to trace hot components of cyanopolyynes.

In this paper, we present ALMA Band 3 data toward five HMPOs taken in Cycle 6.
We focus not only on oxygen(O)-bearing and nitrogen(N)-bearing COMs, but also on carbon-chain species, especially cyanopolyynes (HC$_{2n+1}$N, $n=1,2,3,...$).
We will investigate relationships among these different types of species, and discuss connections between chemical features and physical conditions of the target sources.

The structure of this paper is as follows.
We describe archival data sets and reduction procedure in Section \ref{sec:data}.
In Sections \ref{sec:continuum} and \ref{sec:mom0}, continuum images and moment 0 maps of the observed molecular species are presented.
We present spectral analyses in Section \ref{sec:spectralana}.
We compare spatial distributions among different types of molecules to investigate the chemical differentiation among each source in Section \ref{sec:d1}.
Chemical composition is compared among each core and relationships between the chemistry and physical conditions are discussed in Section \ref{sec:d2}.
Comparisons of chemical composition between observational results and chemical simulations will be discussed in Section \ref{sec:d3}.
In Section \ref{sec:con}, main conclusions of this paper are summarized.

\section{Data Reduction} \label{sec:data}

We have analyzed the ALMA Band 3 data toward five HMPOs\footnote{Proposal ID: 2018.1.00424.S, PI: Caroline Gieser}. 
Details about the five target sources and molecular lines presented in this paper are summarized in Tables \ref{tab:source} and \ref{tab:mol}, respectively.
The data sets consist of six scheduling blocks.
Table \ref{tab:dataset} summarizes information on each data set.
These data sets were obtained with three different frequency setups, corresponding to each column of Table \ref{tab:dataset}.
The data toward G9.62+0.19, G10.47+0.03 and G12.89+0.49 are contained in the same scheduling blocks (the upper rows of Table \ref{tab:dataset}), and the other two sources, G16.57-0.05 and G19.88-0.53, are involved in the other scheduling blocks (the lower rows of Table \ref{tab:dataset}).
The observations of each scheduling block were conducted with different baselines because of the different observing dates.
As a result, the angular resolutions (Ang. Res.) are different among sources and frequency setups.
The bandwidth and frequency resolution of the spectral windows for the molecular lines are 117 MHz and 244 kHz, respectively.
The frequency resolution corresponds to the velocity resolution of $\sim0.8$ km\,s$^{-1}$ at the observed frequency.

We conducted data reduction and imaging using the Common Astronomy Software Application \citep[CASA;][]{2007ASPC..376..127M} on the pipeline-calibrated visibilities.
We ran the calibration scripts using CASA version 5.4.0 and 5.6.1, according to QA2 reports.
The data cubes were created by the {\it {tclean}} task in CASA, combining the data of the 12-m and 7-m arrays.
Briggs weighting with a robust parameter of 0.5 was applied.
The phase reference centers of each source are summarized in Table \ref{tab:source}.

Continuum images ($\lambda=2.75$ mm) were created from the broadest spectral window (the central frequency of 109.5 GHz and the bandwidth of 1.875 GHz) using the {\it {imcontsub}} task in CASA. 
We determined line-free channels checking the {\it {tclean}} images in the {\it {imcontsub}} task, because molecular lines have been detected at the continuum core positions.

\begin{deluxetable*}{lllcccccc}
\tablecaption{Summary of Target Sources \label{tab:source}}
\tablewidth{0pt}
\tablehead{
\colhead{Source} & \colhead{R.A.(J2000)} & \colhead{Decl.(J2000)} & \colhead{$D$\tablenotemark{a}} & \colhead{$V_{\rm {lsr}}$} & \colhead{$L_{\rm {bol}}$\tablenotemark{a}} & \colhead{$M_{\rm {clump}}$\tablenotemark{a}} & \colhead{$N$(H$_{2}$)\tablenotemark{a}} & \colhead{$T_{\rm {dust}}$\tablenotemark{a}} \\
\colhead{} & \colhead{} & \colhead{} & \colhead{(kpc)} & \colhead{(km\,s$^{-1}$)} & \colhead{($L_{\sun}$)} & \colhead{($M_{\sun}$)} & \colhead{(cm$^{-2}$)} & \colhead{(K)}
}
\startdata
G9.62+0.19 & 18$^{\rm {h}}$06$^{\rm {m}}$14\fs9 & -20\degr31\arcmin39\farcs2 & 5.2 & 4.4 & $2.39 \times 10^{5}$ & $3.33 \times10^{3}$ & $1.64 \times 10^{23}$ &  32 \\
G10.47+0.03 & 18$^{\rm {h}}$08$^{\rm {m}}$38\fs2 & -19\degr51\arcmin50\farcs1 & 10.7 & 67.8 & $4.48 \times 10^{5}$ & $2.58 \times 10^{4}$ & $6.35 \times 10^{23}$ & 25 \\
G12.89+0.49 & 18$^{\rm {h}}$11$^{\rm {m}}$51\fs5 & -17\degr31\arcmin28\farcs9 & 3.0 & 33.8 & $1.79 \times 10^{4}$ & $1.26 \times 10^{3}$ & $1.85 \times 10^{23}$ & 23 \\
G16.57-0.05 & 18$^{\rm {h}}$21$^{\rm {m}}$09\fs2 & -14\degr31\arcmin45\farcs5 & 4.7 & 59.1 & $1.82 \times 10^{4}$ & $1.34 \times 10^{3}$ & $1.05 \times 10^{23}$ & 25 \\
G19.88-0.53 & 18$^{\rm {h}}$29$^{\rm {m}}$14\fs7 & -11\degr50\arcmin24\farcs0 & 3.3 & 43.6 & $7.89 \times 10^{3}$ & $1.64 \times 10^{3}$ & $1.82 \times 10^{23}$ & 20 \\
\enddata
\tablenotetext{a}{Taken from \citet{2018MNRAS.473.1059U}.}
\end{deluxetable*}

\begin{deluxetable*}{llcc}
\tabletypesize{\footnotesize}
\tablecaption{Summary of Molecular Lines for Moment 0 Maps\label{tab:mol}}
\tablewidth{0pt}
\tablehead{
\colhead{Species} & \colhead{Transition} & \colhead{Frequency$^\tablenotemark{a}$} & \colhead{$E_{\rm {up}}/k$} \\
\colhead{} & \colhead{} &  \colhead{(GHz)} &  \colhead{(K)} 
}
\startdata
SiO & $2-1$ & 86.84696 & 6.3 \\
Ethynyl (CCH) & $N=1-0, J=\frac{3}{2}-\frac{1}{2}, F=2-1$ & 87.316925 & 4.2  \\
Isocyanic acid (HNCO) & $4_{0,4}-3_{0,3}$ & 87.925237 & 10.5  \\
Cyanoacetylene (HC$_{3}$N) & $10-9$ & 90.979023 & 24.0  \\
Methyl Cyanide (CH$_{3}$CN) & $5_{3}-4_{3}$ & 91.9711307 & 77.5  \\
Vinyl Cyanide (CH$_{2}$CHCN) & $10_{1,10}- 9_{1, 9}$ & 92.42625 & 26.6  \\
Cyanodiacetylene (HC$_{5}$N) & $35-34$ & 93.188123 & 80.5  \\
Methanol (CH$_{3}$OH)  & $1_{0,1}-2_{1,2}$, $v_{t}=1$ & 93.196673 & 302.9  \\
Ethyl Cyanide (CH$_{3}$CH$_{2}$CN) & $11_{0,11}-10_{0,10}$ & 96.919762 & 28.1 \\
SO & $3_{2}-2_{1}$ & 99.29987 & 9.2 \\
Dimethyl ether (CH$_{3}$OCH$_{3}$)$^\tablenotemark{b}$ & $4_{1,4}-3_{0,3}$ (EA, AE, EE, AA) & 99.324362--99.326072  & 10.2  \\
\enddata
\tablenotetext{a}{Rest frequencies were taken from the Cologne Database for Molecular Spectroscopy \citep[CDMS;][]{2005JMoSt.742..215M}. Papers of spectroscopic laboratory experiments are listed in the CDMS website.}
\tablenotetext{b}{Four lines are blended.}
\end{deluxetable*}

\begin{deluxetable*}{lccc}
\tablecaption{Information on data sets presented in this paper \label{tab:dataset}}
\tablewidth{0pt}
\tablehead{
\colhead{} & \colhead{1st setup} &  \colhead{2nd setup} & \colhead{3rd setup}
}
\startdata
Frequency range & 96.06--110.43 GHz & 86.28--100.79 GHz & 90.20--105.61 GHz \\
Contained lines & Continuum & CCH, CH$_{3}$OCH$_{3}$, HNCO, & HC$_{5}$N, HC$_{3}$N, CH$_{3}$OH \\ 
			   &  		& CH$_{3}$CH$_{2}$CN, SiO, SO  & CH$_{3}$CN, CH$_{2}$CHCN \\			   
\cline{1-4}
\multicolumn{4}{l}{\bf {G9.62+0.19, G10.47+0.03, G12.89+0.49}} \\
12-m array observations & 2019 May 3 & 2018 Nov 20 & 2019 Jan 10 \\
Projected baseline distance & 12.0 m--673.0 m & 15.1 m -- 1.4 km & 12.0 m -- 269.2 m \\
Typical Ang. Res. & $1.8\arcsec \times 1.0\arcsec$ & $0.8\arcsec \times 0.8\arcsec$ & $4.5\arcsec \times 2.5\arcsec$ \\
7-m array observations & 2019 Jan 26, Mar 29 & 2019 May 10, 20, 23 & 2019 May 20, 23 \\
\cline{1-4}
\multicolumn{4}{l}{\bf {G16.57-0.05, G19.88-0.53}} \\
12-m array observations & 2019 Apr 17 & 2018 Nov 16, 19 & 2018 Oct 30 \\
Projected baseline distance & 12.0 m -- 782.9 m & 12.6 m -- 1.4 km & 12.5 m -- 1.4 km \\
Typical Ang. Res. & $1.5\arcsec \times 1.0\arcsec$ & $1.0\arcsec \times 0.7\arcsec$ & $1.0\arcsec \times 0.7\arcsec$ \\
7-m array observations & 2019 Apr 1, 17 & 2019 Jan 16, 17 & 2019 Jan 16, 19 \\
\enddata
\end{deluxetable*}

\section{Results and Analyses} \label{sec:res}
\subsection{Continuum images} \label{sec:continuum}

\begin{figure*}[!th]
 \begin{center}
  \includegraphics[bb =0 60 520 792, scale=0.75]{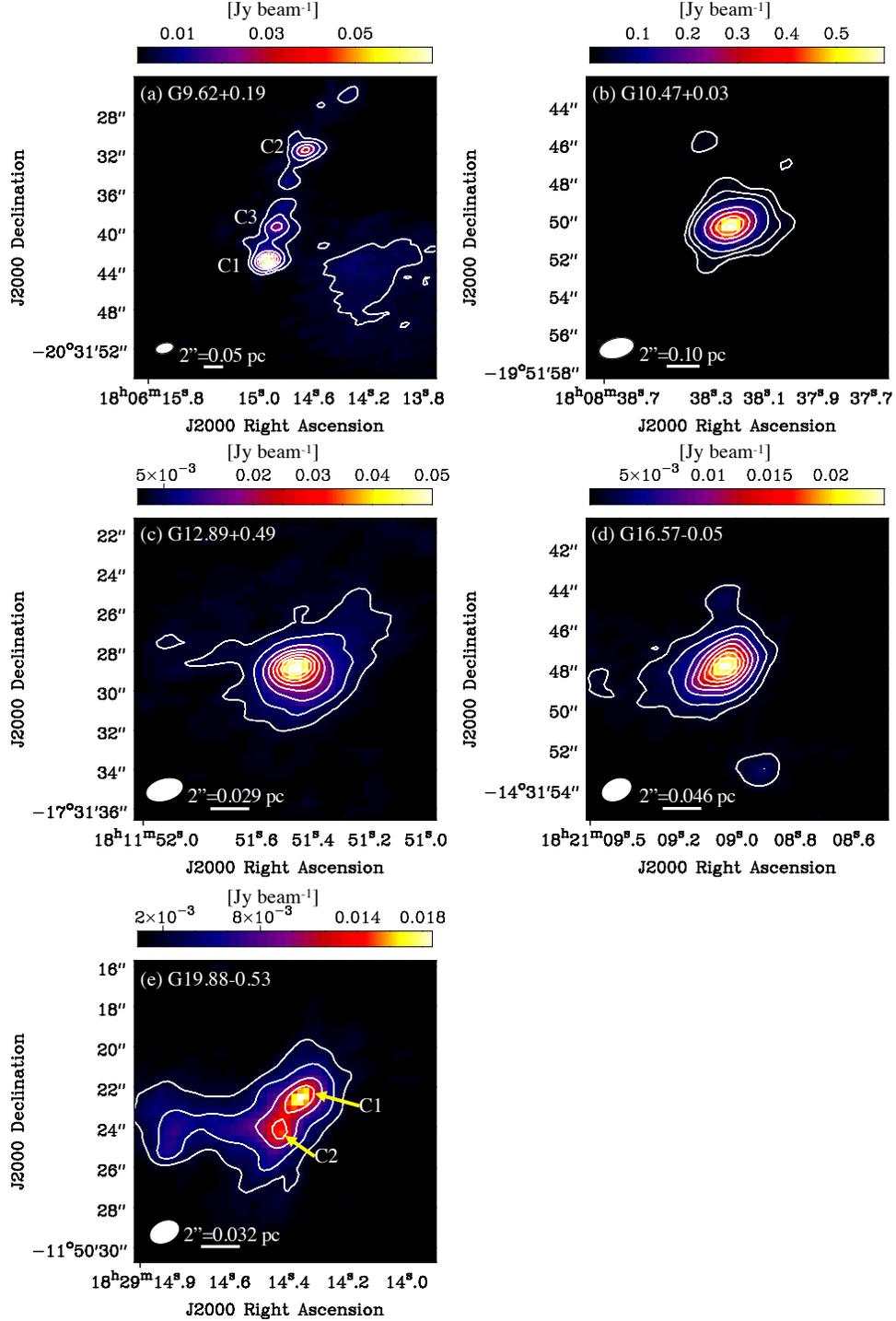}
 \end{center}
\caption{Continuum images ($\lambda=2.75$ mm, combination of the 12-m and 7-m arrays) toward five sources. The filled white ellipse indicates the angular resolution. Information about noise level, contour level, and angular resolution for each panel is summarized in Table \ref{tab:contmap}.  \label{fig:cont}}
\end{figure*}

Figure \ref{fig:cont} shows continuum images ($\lambda=2.75$ mm) toward the five sources.
The angular resolution (Ang. Res.),  beam position angle (PA), rms noise level, and contour levels of each continuum image are summarized in Table \ref{tab:contmap}.
We applied 2D gaussian fitting for the continuum images in CASA, and identified cores.
Table \ref{tab:2Dgauss} summarizes properties of the identified cores.
Three and two cores are identified in G9.62+0.19 and G19.88-0.53, respectively. 
We named them C1, C2, and C3 in the order of peak flux.

In the G9.62+0.19 region, twelve millimeter cores were identified along a filamentary structure in the 1.3 mm continuum emission \citep{2017ApJ...849...25L}.
MM1 is located outside of our image.
Peaks of Cores C1, C2, and C3 we identified correspond to peak positions of cores MM11, MM4, and MM8, respectively, identified by \citet{2017ApJ...849...25L}.
The evolutionary stages of these cores (MM11, MM4, and MM8) were proposed to be ultracompact \ion{H}{2} region (UC \ion{H}{2}), late hot molecular core (HMC) or hyper-compact \ion{H}{2} region (HC \ion{H}{2}), and HMC, respectively \citep{2017ApJ...849...25L}.

The morphology of the continuum emission in G10.47+0.03 is different from ALMA Band 4 data drawn at  angular resolutions of $\sim 0.7\arcsec$ \citep{2020ApJ...895...86G}.
The difference seems to be caused by the beam effect, and the core is unlikely spatially resolved sufficiently, because of the farthest distance (10.7 kpc; Table \ref{tab:source}).
The weak emission at the northeast position corresponds to a sharp feature in the 1.88 mm and 2.30 mm continuum images in Figure 1 in \citet{2020ApJ...895...86G}.

The continuum emission of G12.89+0.49 (also known as IRAS\,18089-1732) shows an elongated structure from northwest to southeast.
This feature is also seen in the 1.2 mm continuum emission \citep{2021ApJ...915L..10S}.
We identify one core in our data.
Even in higher angular resolution data (0.3\arcsec, corresponding to 700 au), there can be seen only one core \citep{2021ApJ...915L..10S}.
The morphology of the continuum emission in G16.57-0.05 (also known as IRAS\,18182-1433) is consistent with a previous result reported by \citet{2022MNRAS.511.3463Q}.

G19.88-0.53 consists of two continuum cores, and the obtained morphology of the continuum image is consistent with that reported by \citet{2020MNRAS.497.5454I}.
Cores C1 and C2 correspond to MM2 and MM1, respectively, identified by \citet{2020MNRAS.497.5454I}.
Both of the cores are associated with the radio emission \citep{2006AJ....131..939Z}, and the 44 GHz CH$_{3}$OH maser has been detected at C2 \citep{2017ApJS..233....4R}.
\citet{2006AJ....131..939Z} suggested that this region is a triple stellar system based on the VLA 1.3 cm and 7 mm continuum images.
Cores C1 and C2 correspond to IRAS\,18264-1152b and IRAS\,18264-1152c respectively \citep{2006AJ....131..939Z}, but we cannot recognize the third one (IRAS\,18264-1152a), which is located at $\sim2\arcsec$ to the west from the two continuum cores.

\begin{deluxetable*}{lcccc}
\tablecaption{Information on the Continuum Maps \label{tab:contmap}}
\tablewidth{0pt}
\tablehead{
\colhead{Panel} & \colhead{Ang. Res.} & \colhead{Beam PA} & \colhead{RMS$^\tablenotemark{a}$} & \colhead{Contour level}
}
\startdata
(a) G9.62+0.19 & 1\farcs85 $\times$ 1\farcs04 & -75.9\degr & 1.0 & 5, 10--60$\sigma$ ($10\sigma$-step) \\
(b) G10.47+0.03 & 1\farcs86 $\times$ 1\farcs04 & -75.5\degr & 3.0 & 5, 10, 20--180$\sigma$ ($40\sigma$-step) \\
(c) G12.89+0.49 & 1\farcs82 $\times$ 1\farcs04 & -74.3\degr & 0.6 &5, 10--80$\sigma$ ($10\sigma$-step) \\
(d) G16.57-0.05 & 1\farcs47 $\times$ 1\farcs00 & -65.3\degr & 0.3 & 5, 10--80$\sigma$ ($10\sigma$-step) \\
(e) G19.88-0.53 & 1\farcs46 $\times$ 1\farcs00 & -64.0\degr & 0.45 & 5, 10--40$\sigma$ ($10\sigma$-step)\\
\enddata
\tablenotetext{a}{Unit is mJy\,beam$^{-1}$.}
\end{deluxetable*}

\begin{deluxetable}{ccccccc}
\rotate
\tabletypesize{\scriptsize}
\tablecaption{Results of 2D Gaussian Fitting \label{tab:2Dgauss}}
\tablewidth{0pt}
\tablehead{
\colhead{Core} & \multicolumn{2}{c}{Peak Position} & \colhead{Size} & \colhead{Size\tablenotemark{a}} & \colhead{PA} & \colhead{Peak Flux} \\
\colhead{} & \colhead{R.A.(J2000)} & \colhead{Decl.(J2000)} & \colhead{(\arcsec)} & \colhead{(pc)} & \colhead{(\degr)} & \colhead{(mJy\,beam$^{-1}$)}
}
\startdata
G9.62+0.19 C1 & 18$^{\rm {h}}$06$^{\rm {m}}$14\fs9297 (0\fs0036) & -20\degr31\arcmin42\farcs9012 (0\farcs0303) & 1.09 (0.22) $\times$ 1.01 (0.34) & 0.027 (0.005) $\times$ 0.025 (0.009) & 9.6 (72.6) & 62.3 (3.1) \\
G9.62+0.19 C2 & 18$^{\rm {h}}$06$^{\rm {m}}$14\fs6557 (0\fs0078) & -20\degr31\arcmin31\farcs6473 (0\farcs0621) & 1.64 (0.41) $\times$ 1.24 (0.32) & 0.041 (0.010) $\times$ 0.031 (0.008) & 110 (80) & 29.0 (2.7) \\
G9.62+0.19 C3 & 18$^{\rm {h}}$06$^{\rm {m}}$14\fs8640 (0\fs0085) & -20\degr31\arcmin39\farcs5432 (0\farcs1168) & 2.60 (0.50) $\times$ 1.93 (0.52) & 0.066 (0.013) $\times$ 0.049 (0.013) & 154 (37) & 19.6 (2.0) \\
G10.47+0.03    & 18$^{\rm {h}}$08$^{\rm {m}}$38\fs2249 (0\fs0007) & -19\degr51\arcmin50\farcs3098 (0\farcs0052) & 1.14 (0.05) $\times$ 0.93 (0.03) & 0.059 (0.003) $\times$ 0.048 (0.001) & 117.3 (9.1) & 588.2 (5.3) \\
G12.89+0.49    & 18$^{\rm {h}}$11$^{\rm {m}}$51\fs4509 (0\fs0057) & -17\degr31\arcmin28\farcs9450 (0\farcs0552) & 2.19 (0.29) $\times$ 1.84 (0.25) & 0.032 (0.004) $\times$ 0.027 (0.004) & 88 (81) & 41.4 (2.6) \\
G16.57-0.05     & 18$^{\rm {h}}$21$^{\rm {m}}$09\fs0448 (0\fs0029) & -14\degr31\arcmin47\farcs9005 (0\farcs0337) & 2.51 (0.13) $\times$ 1.63 (0.08) & 0.057 (0.003) $\times$ 0.037 (0.002) & 125.8 (5.0) & 22.35 (0.78) \\
G19.88-0.53 C1& 18$^{\rm {h}}$29$^{\rm {m}}$14\fs3582 (0\fs0023) & -11\degr50\arcmin22\farcs5733 (0\farcs0295) & 2.74 (0.12) $\times$ 1.25 (0.06) & 0.044 (0.002) $\times$ 0.020 (0.001) & 132 (2) & 17.79 (0.49) \\
G19.88-0.53 C2& 18$^{\rm {h}}$29$^{\rm {m}}$14\fs4285 (0\fs0023) & -11\degr50\arcmin24\farcs0739 (0\farcs0293) & 2.08 (0.14) $\times$ 1.77 (0.15) & 0.033 (0.002) $\times$ 0.028 (0.002) & 56 (22) & 14.83 (0.48) \\
\enddata
\tablecomments{Numbers in parentheses indicate the fitting errors. The reported size and position angle are values deconvolved from beam.}
\tablenotetext{a}{The source distances summarized in Table \ref{tab:source} are adopted.}
\end{deluxetable}

\subsection{Moment 0 maps of molecular lines} \label{sec:mom0}

\begin{figure*}[!th]
 \begin{center}
  \includegraphics[bb =0 20 448 629, scale=1.0]{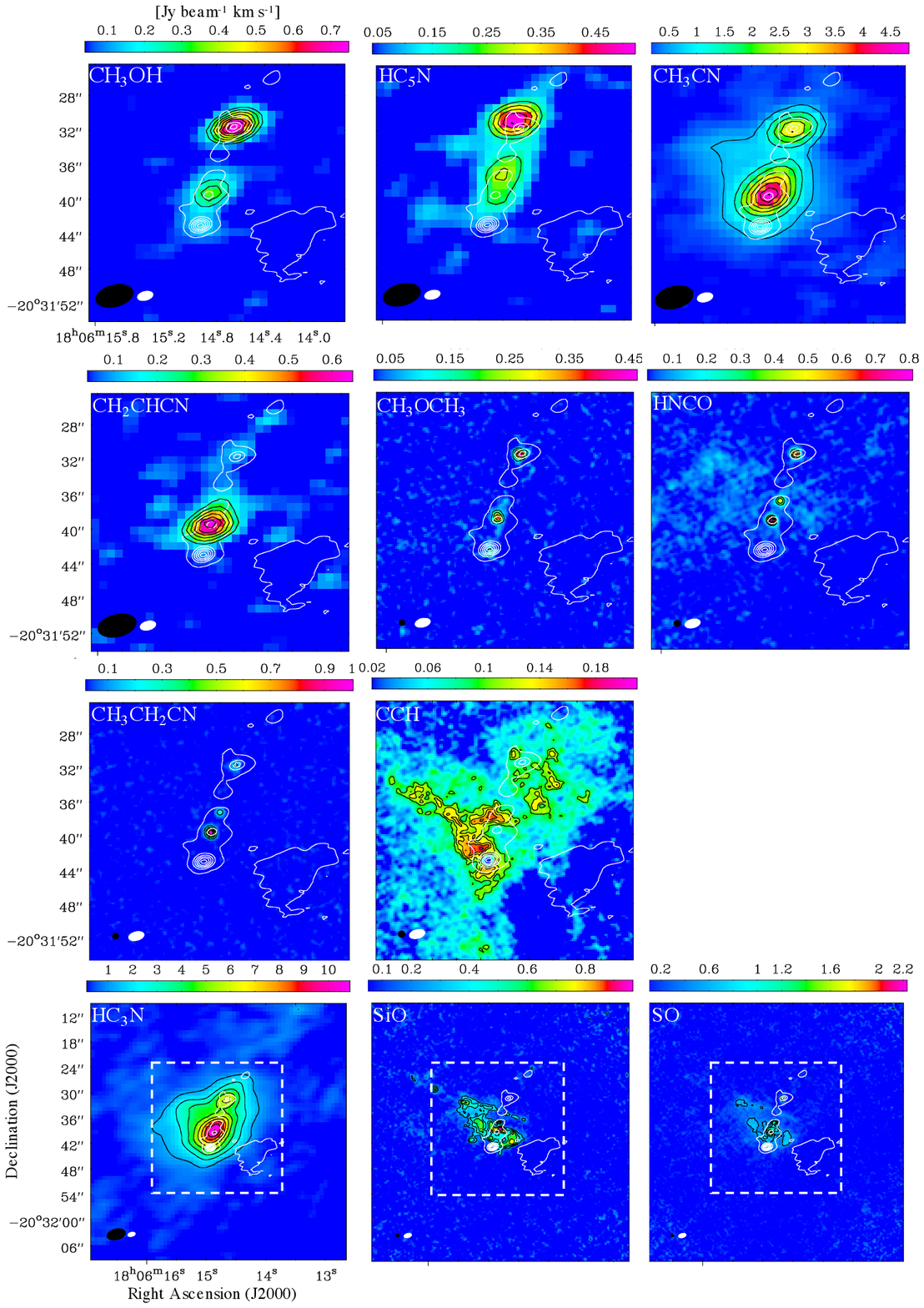}
 \end{center}
\caption{Moment 0 maps (combination of the 12-m and 7-m arrays) of molecular lines toward G9.62+0.19. 
The white contours indicate the continuum emission which is the same one as Figure \ref{fig:cont}. 
The black and white ellipses at the bottom left corner indicate the angular resolution for the moment 0 maps of molecular lines and the continuum emission, respectively.
Information of angular resolution, noise levels, and black contour levels are summarized in Table \ref{tab:mom0map}. 
The dashed squares in the panels of HC$_{3}$N, SiO, and SO indicate the region in the other panels. \label{fig:mom0_G9}}
\end{figure*}

\begin{figure*}[!th]
 \begin{center}
  \includegraphics[bb =0 20 438 624, scale=1.0]{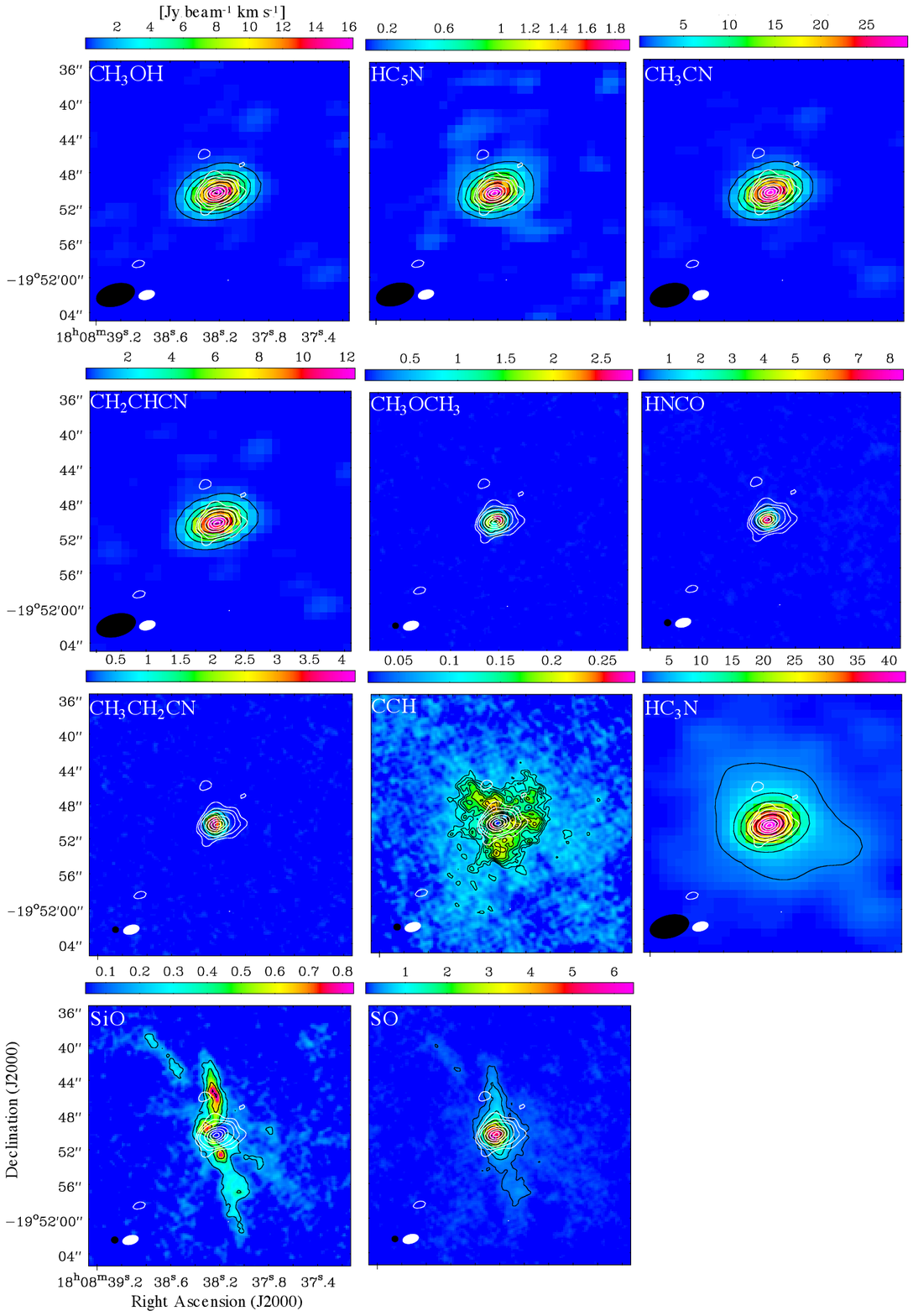}
 \end{center}
\caption{Moment 0 maps (combination of the 12-m and 7-m arrays) of molecular lines toward G10.47+0.03. 
The white contours indicate the continuum emission which is the same one as Figure \ref{fig:cont}. 
The black and white ellipses at the bottom left corner indicate the angular resolution for the moment 0 maps of molecular lines and the continuum emission, respectively.
Information of angular resolution, noise levels, and black contour levels are summarized in Table \ref{tab:mom0map}.  \label{fig:mom0_G10}}
\end{figure*}

\begin{figure*}[!th]
 \begin{center}
  \includegraphics[bb =0 20 440 633, scale=1.0]{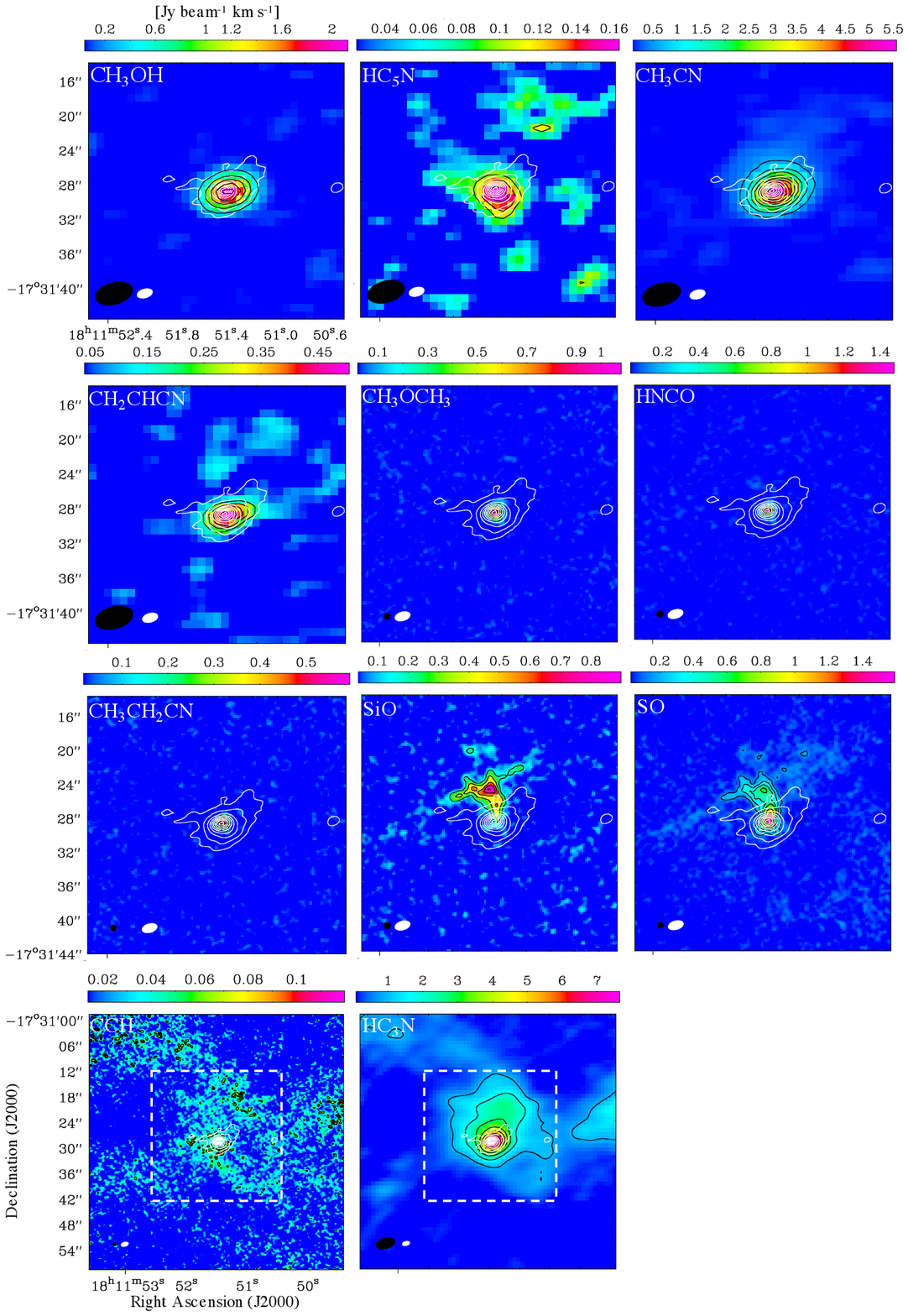}
 \end{center}
\caption{Moment 0 maps (combination of the 12-m and 7-m arrays) of molecular lines toward G12.89+0.49. 
The white contours indicate the continuum emission which is the same one as Figure \ref{fig:cont}. 
The black and white ellipses at the bottom left corner indicate the angular resolution for the moment 0 maps of molecular lines and the continuum emission, respectively.
Information of angular resolution, noise levels, and black contour levels are summarized in Table \ref{tab:mom0map}. 
The dashed squares in the panels of CCH and HC$_{3}$N indicate the region in the other panels. \label{fig:mom0_G12}}
\end{figure*}

\begin{figure*}[!th]
 \begin{center}
  \includegraphics[bb =0 20 464 645, scale=1.0]{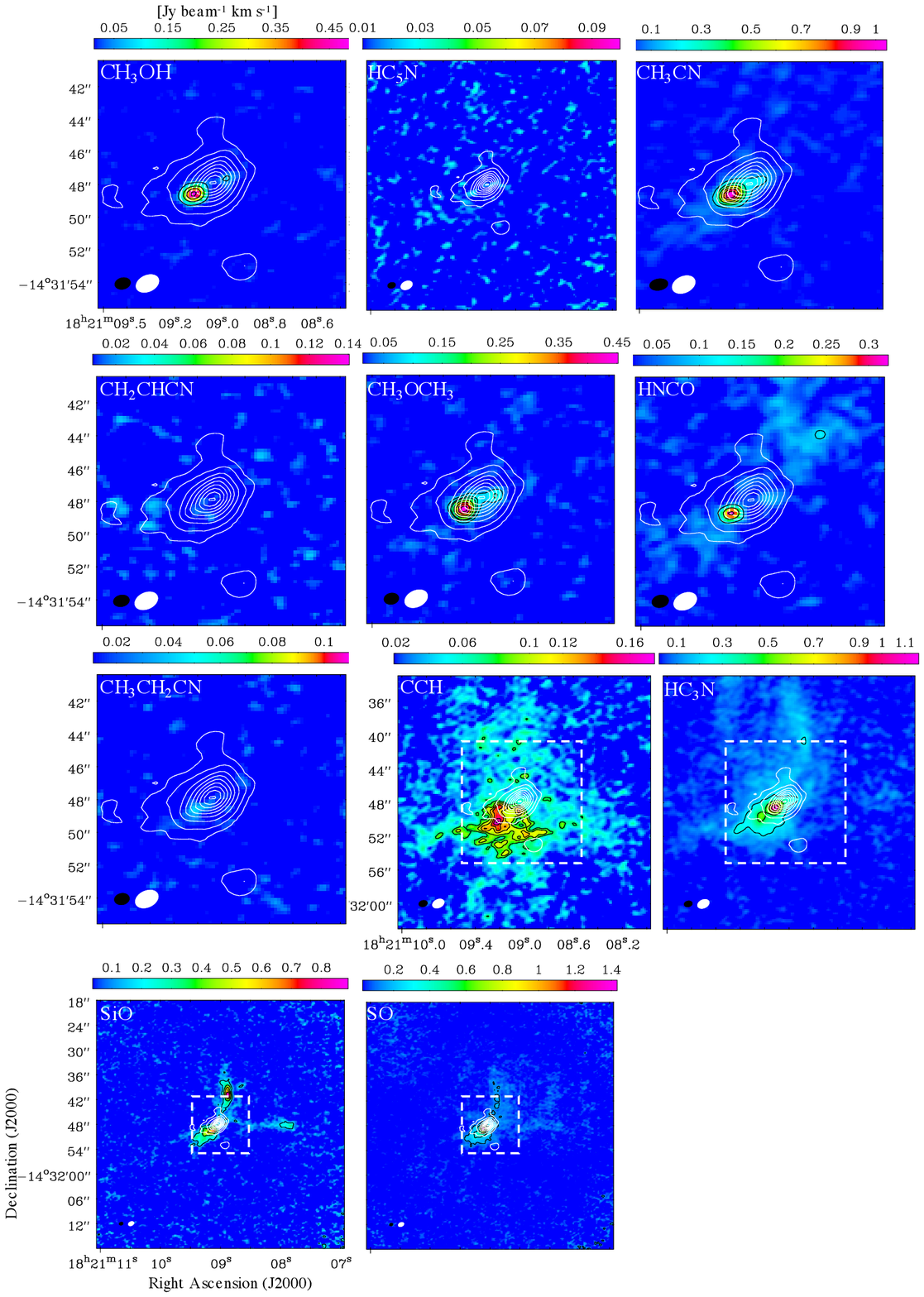}
 \end{center}
\caption{Moment 0 maps (combination of the 12-m and 7-m arrays) of molecular lines toward G16.57-0.05. 
The white contours indicate the continuum emission which is the same one as Figure \ref{fig:cont}. 
The black and white ellipses at the bottom left corner indicate the angular resolution for the moment 0 maps of molecular lines and the continuum emission, respectively.
Information of angular resolution, noise levels, and black contour levels are summarized in Table \ref{tab:mom0map}. 
The dashed squares in the panels of CCH, HC$_{3}$N, SiO and SO indicate the region in the other panels. \label{fig:mom0_G16}}
\end{figure*}

\begin{figure*}[!th]
 \begin{center}
  \includegraphics[bb =0 20 425 609, scale=1.0]{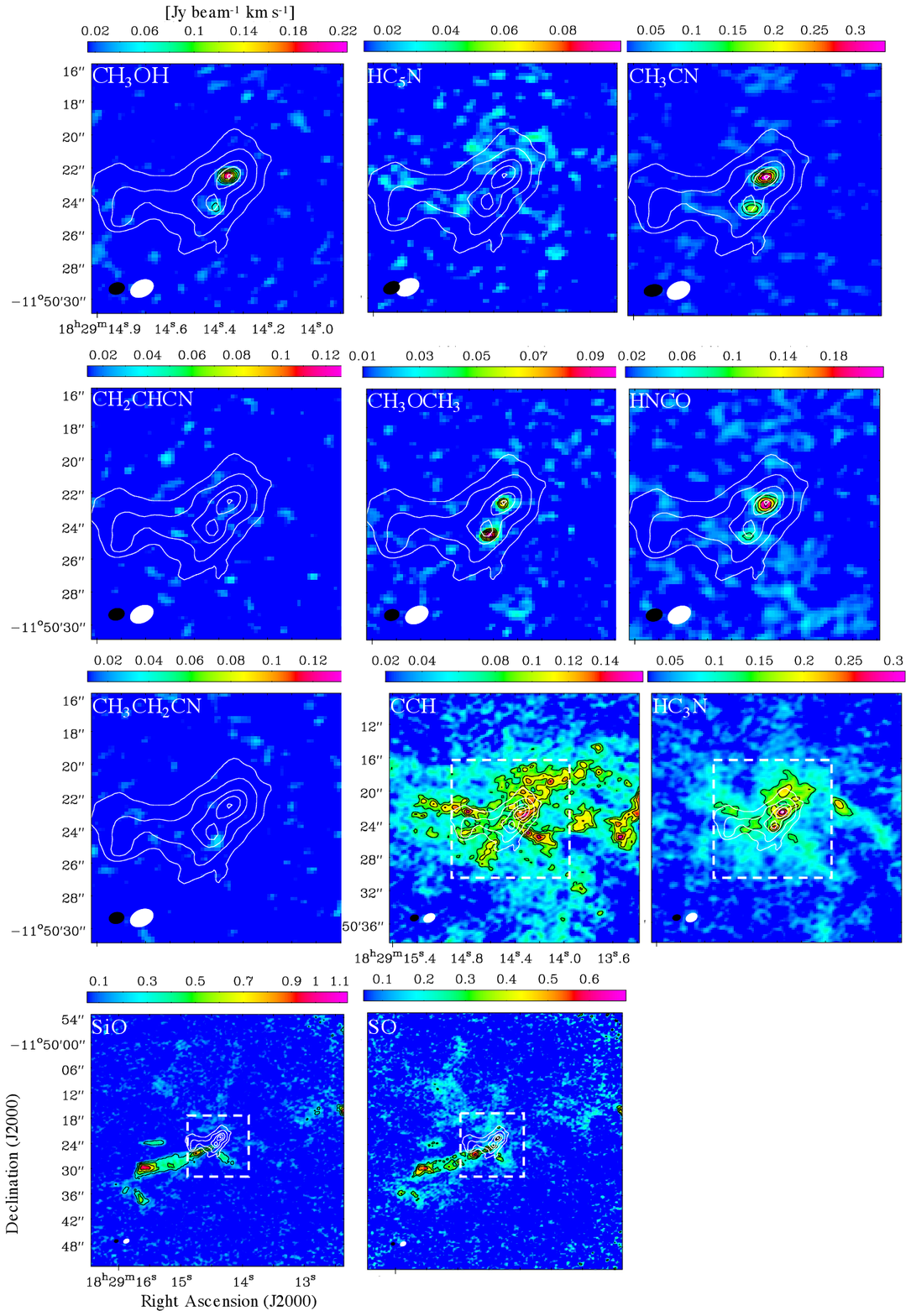}
 \end{center}
\caption{Moment 0 maps (combination of the 12-m and 7-m arrays) of molecular lines toward G19.88-0.53. 
The white contours indicate the continuum emission which is the same one as Figure \ref{fig:cont}. 
The black and white ellipses at the bottom left corner indicate the angular resolution for the moment 0 maps of molecular lines and the continuum emission, respectively.
Information of angular resolution, noise levels, and black contour levels are summarized in Table \ref{tab:mom0map}. 
The dashed squares in the panels of CCH, HC$_{3}$N, SiO and SO indicate the region in the other panels. \label{fig:mom0_G19}}
\end{figure*}

\begin{splitdeluxetable*}{lcccccccBccccccccccccc}
\movetableright=-2in
\rotate
\tabletypesize{\scriptsize}
\tablecaption{Information on the moment 0 maps \label{tab:mom0map}}
\tablewidth{0pt}
\tablehead{
\colhead{Molecule} & \multicolumn{3}{c}{G9.62+0.19} & \colhead{} & \multicolumn{3}{c}{G10.47+0.03} &  \colhead{} & \multicolumn{3}{c}{G12.89+0.49} &  \colhead{} & \multicolumn{3}{c}{G16.57-0.05} &  \colhead{} & \multicolumn{3}{c}{G19.88-0.53} \\
\cline{2-4} \cline{6-8} \cline{10-12} \cline{14-16} \cline{18-20}
\colhead{} & \colhead{Ang. Res. (PA)} & \colhead{RMS$^\tablenotemark{a}$} & \colhead{Contour levels} & \colhead{} & \colhead{Ang. Res. (PA)} & \colhead{RMS$^\tablenotemark{a}$} & \colhead{Contour levels} & \colhead{} & \colhead{Ang. Res. (PA)} & \colhead{RMS$^\tablenotemark{a}$} & \colhead{Contour levels} & \colhead{} & \colhead{Ang. Res. (PA)} & \colhead{RMS$^\tablenotemark{a}$} & \colhead{Contour levels} & \colhead{} & \colhead{Ang. Res. (PA)} & \colhead{RMS$^\tablenotemark{a}$} & \colhead{Contour levels} 
}
\startdata
CH$_{3}$OH & 4\farcs5 $\times$ 2\farcs5 (-75.7\degr) & 0.04 &5--17$\sigma$ ($2\sigma$-step) & & 4\farcs5 $\times$ 2\farcs5 (-74.8\degr) & 0.1 & 10--150$\sigma$ ($20\sigma$-step) & & 4\farcs4 $\times$ 2\farcs5 (-74.5\degr) & 0.04 & 10--50$\sigma$ ($10\sigma$-step) & & 0\farcs97 $\times$ 0\farcs71 (-76.3\degr) & 0.013 & 5--35$\sigma$ ($10\sigma$-step) & &  0\farcs97 $\times$ 0\farcs71 (-75.1\degr) & 0.014 & 4-12$\sigma$ ($2\sigma$-step) \\
HC$_{5}$N & 4\farcs5 $\times$ 2\farcs5 (-75.7\degr) & 0.04 & 5--13$\sigma$ ($2\sigma$-step) & & 4\farcs5 $\times$ 2\farcs5 (-74.8\degr) & 0.05 & 5--35$\sigma$ ($5\sigma$-step) & & 4\farcs4 $\times$ 2\farcs5 (-74.5\degr) & 0.025 & 4,5,6$\sigma$ & & 0\farcs97 $\times$ 0\farcs71 (-76.3\degr) & 0.01 & ... & &  0\farcs97 $\times$ 0\farcs71 (-75.1\degr) & 0.012 & ... \\
CH$_{3}$CN & 4\farcs5 $\times$ 2\farcs6 (-76.4\degr) & 0.06 &10--80$\sigma$ ($10\sigma$-step) & & 4\farcs6 $\times$ 2\farcs6 (-75.2\degr) & 0.2 & 10--130$\sigma$ ($20\sigma$-step) & & 4\farcs5 $\times$ 2\farcs6 (-74.7\degr) & 0.06 & 10--90$\sigma$ ($10\sigma$-step) & & 1\farcs1 $\times$ 0\farcs71 (-80.1\degr) & 0.015 & 10--60$\sigma$ ($10\sigma$-step) & &  1\farcs1 $\times$ 0\farcs72 (-79.6\degr) & 0.017 & 5--17$\sigma$ ($3\sigma$-step) \\
CH$_{2}$CHCN & 4\farcs5 $\times$ 2\farcs6 (-76.7\degr) & 0.04 &5--15$\sigma$ ($2\sigma$-step) & & 4\farcs6 $\times$ 2\farcs6 (-75.4\degr) & 0.09 & 10--130$\sigma$ ($20\sigma$-step) & & 4\farcs5 $\times$ 2\farcs6 (-74.8\degr) & 0.04 & 5--11$\sigma$ ($2\sigma$-step) & & 0\farcs99 $\times$ 0\farcs71 (-78.5\degr) & 0.008 & ... & &  0\farcs97 $\times$ 0\farcs72 (-77.2\degr) & 0.013 & ... \\
CH$_{3}$OCH$_{3}$ & 0\farcs74 $\times$ 0\farcs70 (82.9\degr) & 0.02 &5--20$\sigma$ ($5\sigma$-step) & & 0\farcs74 $\times$ 0\farcs70 (81.3\degr) & 0.03 & 10--90$\sigma$ ($20\sigma$-step) & & 0\farcs73 $\times$ 0\farcs70 (83.0\degr) & 0.017 & 10--60$\sigma$ ($10\sigma$-step) & & 0\farcs92 $\times$ 0\farcs70 (-79.9\degr) & 0.014 & 5,8,10,15,20,25,30$\sigma$ & &  0\farcs91 $\times$ 0\farcs68 (-77.4\degr) & 0.010 & 5--10$\sigma$ \\
HNCO & 0\farcs81 $\times$ 0\farcs76 (84.8\degr) & 0.025 &10--30$\sigma$ ($5\sigma$-step) & & 0\farcs81 $\times$ 0\farcs76 (84.1\degr) & 0.04 & 10, 50--200$\sigma$ ($50\sigma$-step) & & 0\farcs80 $\times$ 0\farcs76 (84.0\degr) & 0.025 & 10--50$\sigma$ ($10\sigma$-step) & & 1\farcs0 $\times$ 0\farcs76 (-76.9\degr) & 0.02 & 5,10,15$\sigma$& &  1\farcs0 $\times$ 0\farcs76 (-75.5\degr) & 0.015 & 5--14$\sigma$ ($3\sigma$-step) \\
CH$_{3}$CH$_{2}$CN & 0\farcs74 $\times$ 0\farcs70 (83.4\degr) & 0.017 & 10--50$\sigma$ ($10\sigma$-step) & & 0\farcs74 $\times$ 0\farcs71 (81.1\degr) & 0.041 & 10,30,50,70,90,100$\sigma$  & & 0\farcs73 $\times$ 0\farcs70 (81.3\degr) & 0.017 & 5--30$\sigma$ ($5\sigma$-step) & & 0\farcs93 $\times$ 0\farcs70 (-79.8\degr) & 0.011 & ... & &  0\farcs92 $\times$ 0\farcs70 (-78.3\degr) & 0.009 & ... \\
CCH & 0\farcs82 $\times$ 0\farcs76 (87.1\degr) & 0.02 &5--9$\sigma$ & & 0\farcs82 $\times$ 0\farcs77 (87.5\degr) & 0.02 & 5--12$\sigma$ & & 0\farcs81 $\times$ 0\farcs76 (87.8\degr) & 0.014 & 4--7$\sigma$ & & 1\farcs0 $\times$ 0\farcs76 (-76.6\degr) & 0.019 & 4--9$\sigma$  & &  1\farcs0 $\times$ 0\farcs76 (-75.1\degr) & 0.02 & 4--7$\sigma$ \\
HC$_{3}$N & 4\farcs5 $\times$ 2\farcs6 (-76.2\degr) & 0.12 &10--90$\sigma$ ($10\sigma$-step) & & 4\farcs6 $\times$ 2\farcs6 (-75.1\degr) & 0.3 & 10--130$\sigma$ ($20\sigma$-step) & & 4\farcs6 $\times$ 2\farcs6 (-74.1\degr) & 0.09 & 10--80$\sigma$ ($10\sigma$-step) & & 1\farcs0 $\times$ 0\farcs72 (-77.7\degr) & 0.023 & 10--50$\sigma$ ($10\sigma$-step) & &  1\farcs0 $\times$ 0\farcs73 (-76.9\degr) & 0.025 & 5,7,9,11$\sigma$ \\
SiO & 0\farcs81 $\times$ 0\farcs76 (87.8\degr) & 0.05 & 5--17$\sigma$ ($3\sigma$-step) & & 0\farcs81 $\times$ 0\farcs76 (87.3\degr) & 0.04 & 5--20$\sigma$ ($5\sigma$-step)  & & 0\farcs80 $\times$ 0\farcs76 (87.2\degr) & 0.035 & 5--25$\sigma$ ($5\sigma$-step) & & 1\farcs0 $\times$ 0\farcs76 (-76.8\degr) & 0.036 & 5--20$\sigma$ ($5\sigma$-step) & &  1\farcs0 $\times$ 0\farcs76 (-75.7\degr) & 0.05 & 5--17$\sigma$ ($4\sigma$-step) \\
SO & 0\farcs74 $\times$ 0\farcs70 (82.6\degr) & 0.05 & 10,15,20,30,40$\sigma$ & & 0\farcs74 $\times$ 0\farcs70 (82.2\degr) & 0.05 & 5,10, 20--120$\sigma$ ($20\sigma$-step)  & & 0\farcs73 $\times$ 0\farcs70 (83.4\degr) & 0.03 & 5,10,15, 20--50$\sigma$ ($10\sigma$-step) & & 0\farcs92 $\times$ 0\farcs70 (-79.9\degr) & 0.03 & 5,10,15,20,30,40$\sigma$ & & 0\farcs90 $\times$ 0\farcs68 (-77.4\degr) & 0.054 & 5,7,9,11$\sigma$ \\
\enddata
\tablenotetext{a}{Unit is Jy\,beam$^{-1}$ km\,s$^{-1}$.}
\end{splitdeluxetable*}

Figures \ref{fig:mom0_G9} -- \ref{fig:mom0_G19} show moment 0 maps of molecular lines toward the five sources (combination of the 12-m and 7-m arrays).
The line's information used in these moment 0 maps are summarized in Table \ref{tab:mol}.
Except for G10.47+0.03 (Figure \ref{fig:mom0_G10}), some panels (CCH, HC$_{3}$N, SiO, and SO) show different scales, because their emission regions are extended.
These panels are presented later.
The white dashed squares in these panels indicate regions of the other maps showing the most zoom-in images presented earlier.
Four lines of CH$_{3}$OCH$_{3}$ are not resolved due to very close frequencies.
Thus, we made moment 0 maps including all of these lines.
Information about angular resolution (Ang. Res.), beam position angle (PA), rms noise level, and contour levels is summarized in Table \ref{tab:mom0map}.
Here we note that these data sets were obtained on different days (see Table \ref{tab:dataset} in Section \ref{sec:data}), and then we will compare the molecular lines observed simultaneously in the following parts of this paper.
For instance, CH$_{3}$OH, HC$_{5}$N, HC$_{3}$N, CH$_{3}$CN, and CH$_{2}$CHCN can be compared straightforwardly in each source.
We confirmed that the peak positions of the continuum emission obtained from the widest frequency band (Figure \ref{fig:cont}) are consistent with those made from data containing molecular lines (bandwidth is 117 MHz) by the {\it {imcontsub}} task.
We thus can discuss consistency and inconsistency between peaks of the molecular lines and continuum emission.

All of the molecular lines have been detected in G9.62+0.19 (Figure \ref{fig:mom0_G9}), G10.47+0.03 (Figure \ref{fig:mom0_G10}), and G12.89+0.49 (Figure \ref{fig:mom0_G12}).
On the other hand, all of the molecular lines except for HC$_{5}$N and N-bearing COMs (CH$_{2}$CHCN and CH$_{3}$CH$_{2}$CN) have been detected from G16.57-0.05 (Figure \ref{fig:mom0_G16}) and G19.88-0.53 (Figure \ref{fig:mom0_G19}).

The peak positions of spatial distributions of the HC$_{5}$N ($J=35-34$; $E_{\rm {up}}/k=80.5$ K) line are coincident with the continuum peaks shown as white contours in G9.62+0.19, G10.47+0.03, and G12.89+0.49.
In G9.62+0.19, the peak of the weaker HC$_{5}$N emission is not completely consistent with C3, but it is consistent with MM7 identified in the 1.3 mm continuum emission \citep{2017ApJ...849...25L}.
The MM7 was suggested to be at the early HMC stage \citep{2017ApJ...849...25L}.
If the evolutionary sequence in this region is true, HC$_{5}$N seems to be enhanced in the HC \ion{H}{2} region or HMC stages.

In general, the spatial distribution of HC$_{3}$N is more extended than that of HC$_{5}$N.
This is caused by the fact that the upper-state energy of the HC$_{3}$N line ($E_{\rm {up}}/k=24.0$ K) is much lower than that of HC$_{5}$N, and could be excited in warm envelopes.
The peak positions of the HC$_{3}$N emission are consistent with the continuum peaks, except for G16.57-0.05 where the HC$_{3}$N peak is slightly shifted in a south-eastward direction from the continuum peak.
The peak shift of molecular emission from the continuum emission is also seen in other COMs, and in a figure presented by \citet{2022MNRAS.511.3463Q}.
Although we could not confirm the origin of the peak shifts with the current angular-resolution data, this source may consist of a binary system with a molecular-line-rich source and a line-poor source.
In G9.62+0.19, the emission peak of HC$_{3}$N is almost consistent with C3, but the peak is also extended to MM7.
From comparison of features of HC$_{5}$N and HC$_{3}$N, HC$_{5}$N is more enhanced at Core C2, which is likely irradiated by the UV radiation from the central source in the HC \ion{H}{2} region.

Spatial distributions of CCH are clearly different from the other molecules.
The spatial distributions of CCH are extended.
Holes can be seen at the continuum peaks in G9.62+0.19, G10.47+0.03, and G12.89+0.49.
Such anti-correlations between CCH and continuum emission, and cyanopolyynes support the prediction of chemical simulations \citep{2019ApJ...881...57T} and interpretation of single-dish observations \citep{2021ApJ...908..100T} (see discussion in Section \ref{sec:d1}).
Around Core C2 of G9.62+0.19, the CCH emission is very fragmented.
This may be caused by the cold envelopes blown away by the stellar feedback in Core C2.

The CH$_{3}$OH, CH$_{3}$OCH$_{3}$ and HNCO lines have been detected from all of the sources, and their peaks of the spatial distributions are consistent with the continuum cores except for G16.57-0.05.
The molecular peaks in G16.57-0.05 are consistent with each other and shifted in a south-eastward direction from the continuum peak as well as the HC$_{3}$N line.
In G9.62+0.19, an HNCO peak is located between cores C2 and C3, and not coincident with the continuum peak.
This HNCO peak is consistent with a dense hot core MM7 associated with molecular outflows identified in \citet{2017ApJ...849...25L}.

The CH$_{3}$CN line has been detected in all of the target sources.
As in the case of other molecular lines in G16.57-0.05, the peak position of CH$_{3}$CN is located at the south-east position from the continuum core.

In G9.62+0.19, the N-bearing COMs show stronger peak fluxes at C3 compared to C2.
This trend is opposite to CH$_{3}$OH, CH$_{3}$OCH$_{3}$, and HC$_{5}$N, but similar to HNCO and HC$_{3}$N.
This may mean that these N-bearing COMs prefer the younger HMC stages, rather than the evolved HC \ion{H}{2} region stage.
In the CH$_{3}$CH$_{2}$CN moment 0 map toward G9.62+0.19, a peak is located between cores C2 and C3, and this peak is coincident with the HNCO peak, or MM7 \citep{2017ApJ...849...25L}.

\clearpage
\subsection{Spectral analyses} \label{sec:spectralana}

Under the LTE assumption, we derived the physical conditions using the CH$_{3}$CN $J=5-4$ $K$-ladder ($K=0-4$) emission ($E_{\rm {up}}/k =$ 13.2, 20.4, 41.8, 77.5, and 127.5 K for $K$ = 0, 1, 2, 3, and 4, respectively).
The upper energy levels, Einstein coefficients, degeneracies, partition functions are from the CDMS database \citep{2016JMoSp.327...95E}.
Details about formalism for the CASSIS software are provided by a document\footnote{\url{http://cassis.irap.omp.eu/docs/RadiativeTransfer.pdf}}.
We measured rms noise levels of spectra, and these noise levels were taken into consideration in the spectral fitting.
The coordinates of each core are summarized in Table \ref{tab:2Dgauss}.

\begin{figure*}[!th]
 \begin{center}
  \includegraphics[bb =18 220 567 805, scale=0.8]{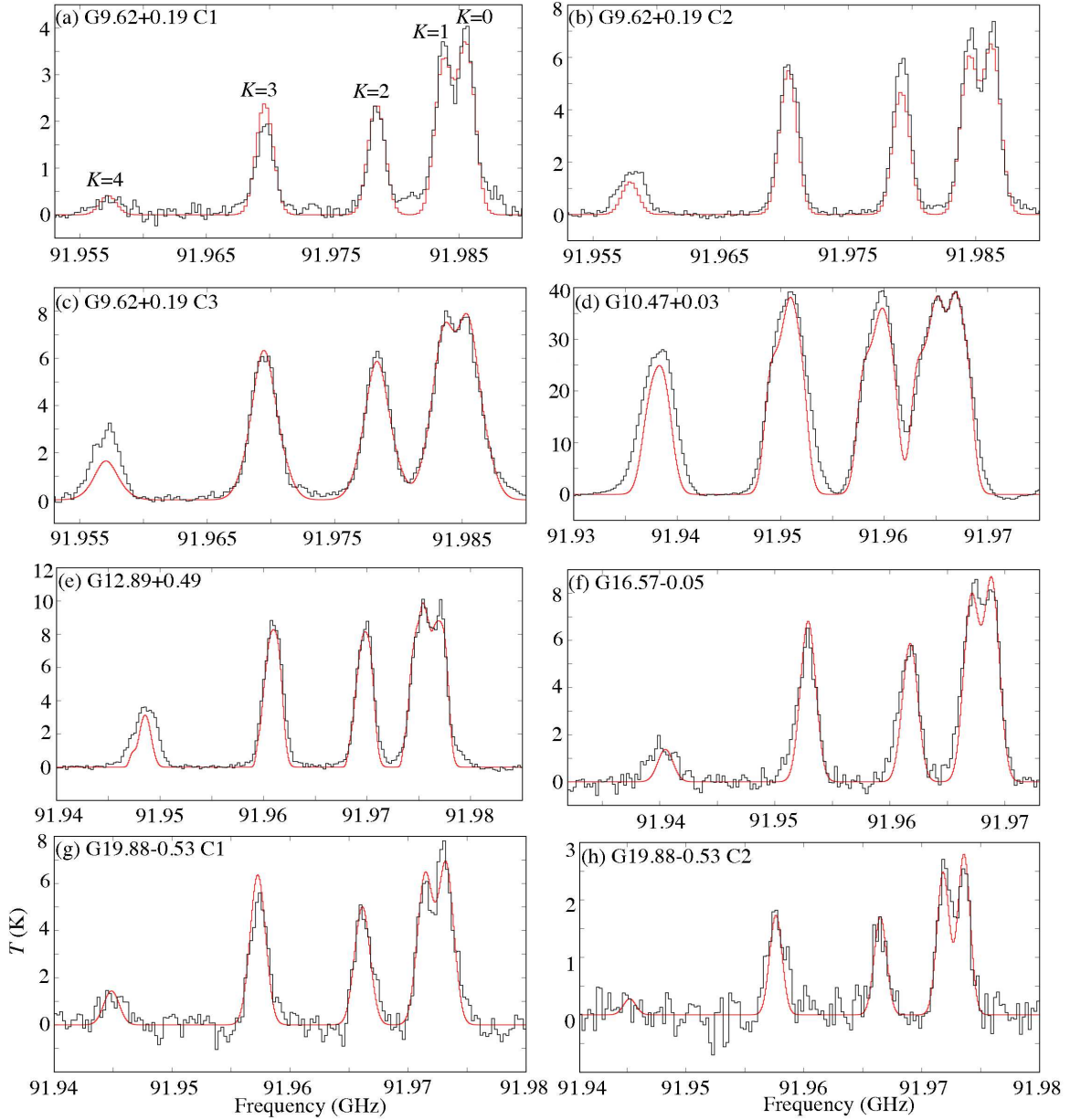}
 \end{center}
\caption{Spectra of the CH$_{3}$CN lines ($J=5_{K}-4_{K}$, $K=0-4$) at each core. The red curves show the best-fit results in the CASSIS software. \label{fig:CH3CN_line}}
\end{figure*}

Figure \ref{fig:CH3CN_line} shows the spectra of CH$_{3}$CN at each core.
Its $K$-ladder lines have been detected from all of the cores.
The column density ($N$), excitation temperature ($T_{\rm {ex}}$), line width (FWHM), systemic velocity ($V_{\rm {LSR}}$), and size were treated as free parameters in the MCMC method for the CH$_{3}$CN lines.
Table \ref{tab:CH3CN_MCMC} summarizes the results of the MCMC analysis.
Red curves in Figure \ref{fig:CH3CN_line} indicate the best-fit models. 
We applied two-component fitting for the spectra at G9.62+0.19 C3, G10.47+0.03, and G12.89+0.49, because the one-component fitting could not reproduce the observed spectra.
The wing emission can be seen, especially in the $K=4$ line.
This wing emission is likely to originate from molecular outflows, because the outflow tracers such as the SiO and SO lines have been detected from all of the sources (Section \ref{sec:mom0}). 
We then excluded the $K=4$ line from the fitting in the MCMC analysis due to non-gaussian line profiles.
Table \ref{tab:tau} in Appendix \ref{sec:a1} summarizes the optical depth of each line.
For sources containing line emission with different optical depths, the column densities and source sizes can be constrained reasonably.
For those with only optically thin line emission, the column densities and source sizes are degenerate. 
However, the derived excitation temperature, line width, and velocity are not or less affected by this.
To compare the abundance ratios with the other molecules, if the source size is fixed, this will not significantly affect the results.

\begin{figure*}[!th]
 \begin{center}
  \includegraphics[bb =10 646 582 802, scale=0.85]{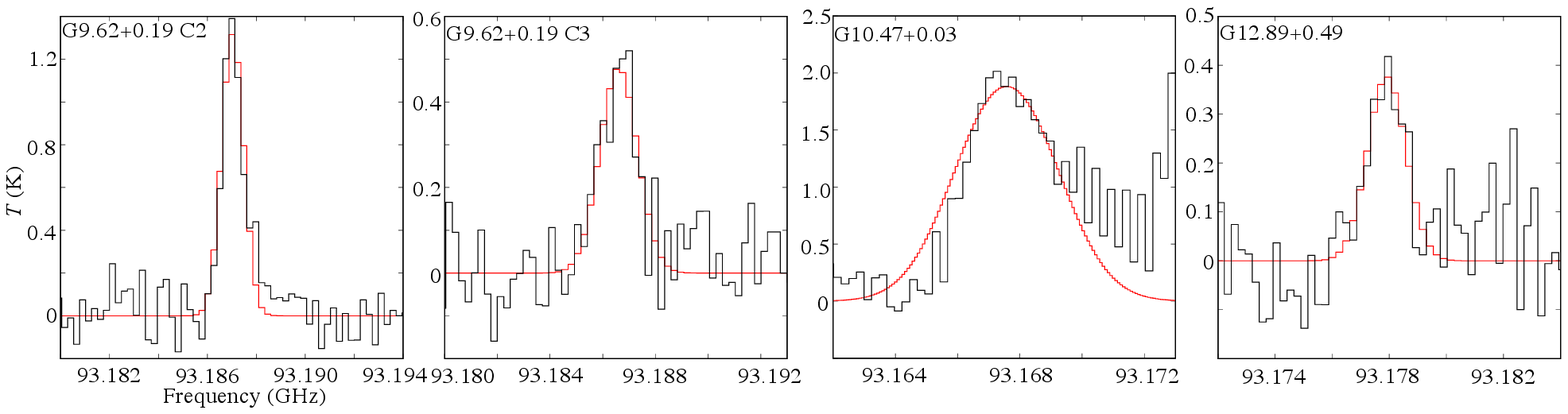}
 \end{center}
\caption{Spectra of the HC$_{5}$N ($J=35-34$) line at each core. The red curves show the best-fit results in the CASSIS software. \label{fig:HC5N_line}}
\end{figure*}

\begin{figure*}[!th]
 \begin{center}
  \includegraphics[bb =0 188 581 806, scale=0.85]{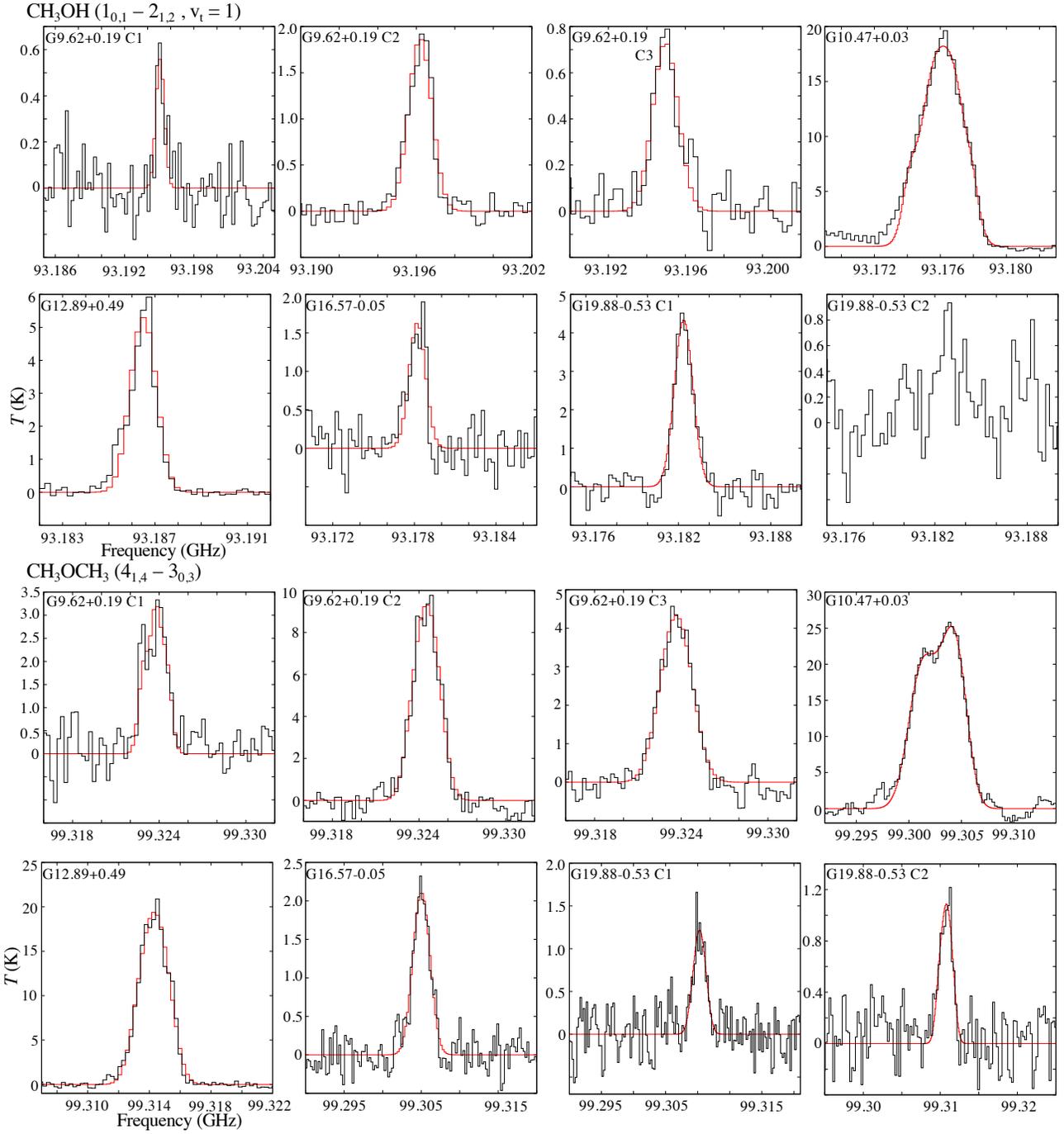}
 \end{center}
\caption{Spectra of the CH$_{3}$OH ($1_{1,0}-2_{1,2}$, $v_{t}=1$) and CH$_{3}$OCH$_{3}$ ($4_{1,4}-3_{0,3}$) lines at each core. The red curves show the best-fit results in the CASSIS software. \label{fig:OCOM_line}}
\end{figure*}

\begin{figure*}[!th]
 \begin{center}
  \includegraphics[bb = 8 501 580 798, scale=0.85]{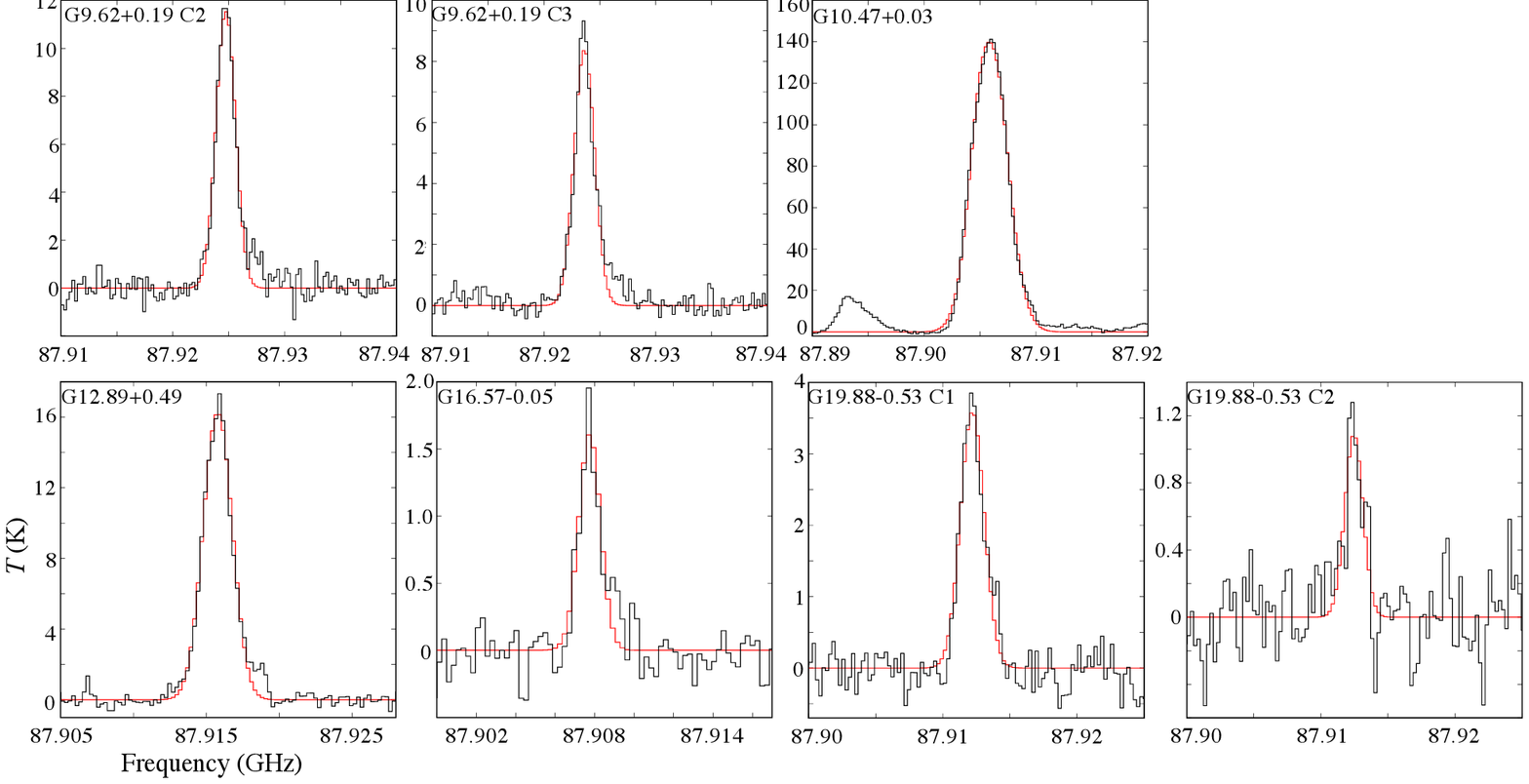}
 \end{center}
\caption{Spectra of the HNCO ($4_{0,4}-3_{0,3}$) line at each core. The red curves show the best-fit results in the CASSIS software. \label{fig:HNCO_line}}
\end{figure*}

\begin{figure*}[!th]
 \begin{center}
  \includegraphics[bb =8 461 583 795, scale=0.85]{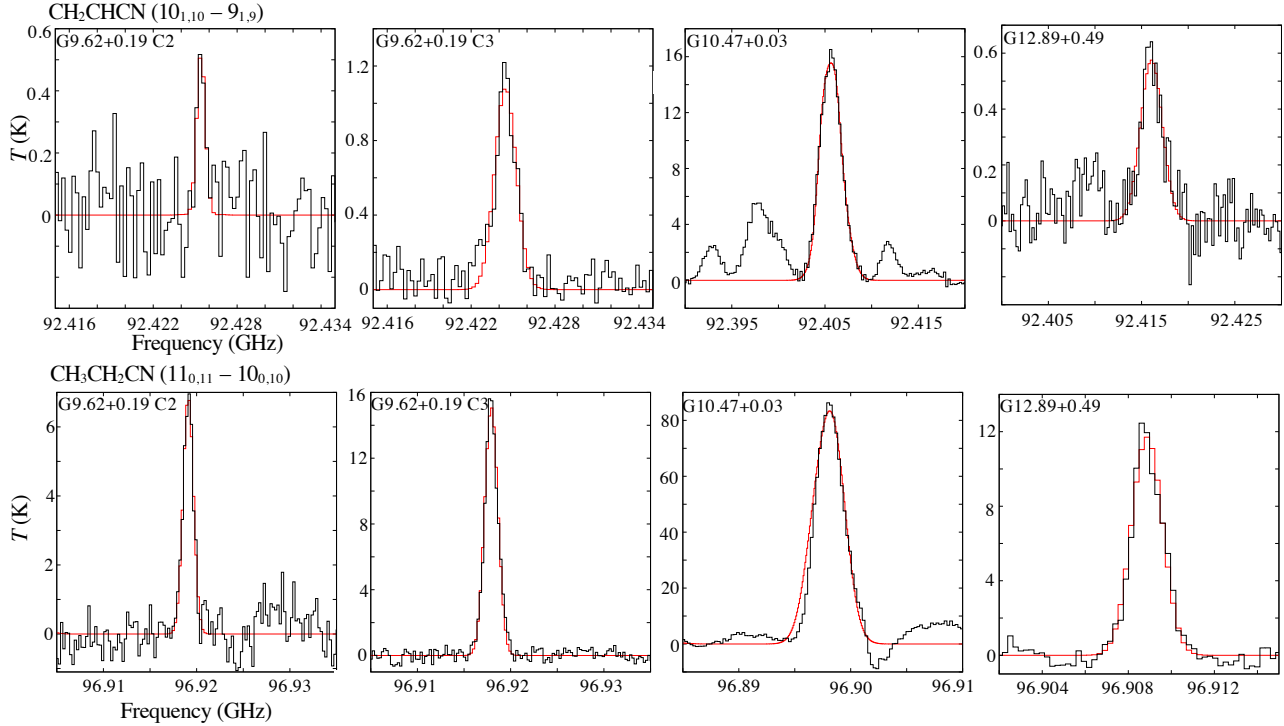}
 \end{center}
\caption{Spectra of the CH$_{2}$CHCN ($10_{1,10}-9_{1,9}$) and CH$_{3}$CH$_{2}$CN ($11_{0,11}-10_{0,10}$) lines at each core. The red curves show the best-fit results in the CASSIS software. \label{fig:NCOM_line}}
\end{figure*}

Figures \ref{fig:HC5N_line}--\ref{fig:NCOM_line} show spectra of HC$_{5}$N, O-bearing COMs (CH$_{3}$OH and CH$_{3}$OCH$_{3}$), HNCO, and N-bearing COMs (CH$_{2}$CHCN and CH$_{3}$CH$_{2}$CN) at each core, respectively.
The HC$_{5}$N ($J=35-34$) line has been detected from four cores; G9.62+0.19 C2, G9.62+0.19 C3, G10.47+0.03, and G12.89+0.49 (Figure \ref{fig:HC5N_line}).
The CH$_{3}$OH line has been detected with sufficient signal-to-noise (S/N) ratios from all of the cores except for G19.88-0.53 C2, in which the line seems to be tentatively detected (Figure \ref{fig:OCOM_line}).
We could not fit the CH$_{3}$OH line at G19.88-0.53 C2.
We note that the upper-state energy of the observed CH$_{3}$OH line (302.9 K) is high.
This may cause the non-detection of the CH$_{3}$OH line in G19.88-0.53 C2, and its non-detection does not mean that CH$_{3}$OH is deficient.

The CH$_{3}$OCH$_{3}$ line has been detected from all of the cores.
The CH$_{3}$OCH$_{3}$ spectrum in G10.47+0.03 shows a clear double-peaked feature.
As seen in its moment 0 map (Figure \ref{fig:mom0_G10}), its spatial distribution is a ring-like structure.
The dust optical thickness at Band 3 is derived to be $\sim 0.07-0.16$, with an assumption of $T_{\rm {dust}}=100-200$ K.
Thus, the dust emission is optically thin at this frequency band.
In addition, the brightness temperature of the continuum emission is around 18 K, which is much lower than the excitation temperatures of molecular lines ($\sim137-195$ K; Table \ref{tab:CH3CN_MCMC}).
Hence, we concluded that such a ring-like feature is not caused by the absorption against the continuum emission.
We constructed channel maps of the CH$_{3}$OCH$_{3}$ line toward this source as shown in Figure \ref{fig:G10_CH3OCH3_channel} in Appendix \ref{sec:a2} to investigate the origin of this unique structure.
A rotating and expanding shell motion plus skewed emission distribution toward the northwest could reproduce the observed velocity structure.
We will discuss detailed analyses of the kinematics of this line in a future paper.

The HNCO line has been detected from all of the cores except for G9.62+0.19 C1 (Figure \ref{fig:HNCO_line}).
The N-bearing COMs have been detected from the four cores (G9.62+0.19 C2, G9.62+0.19 C3, G10.47+0.03, and  G12.89+0.49) as shown in Figure \ref{fig:NCOM_line}, and these four cores match ones where the HC$_{5}$N line has been detected (Figure \ref{fig:HC5N_line}).
The absorption feature at the blue-shifted side in the CH$_{3}$CH$_{2}$CN spectrum in G10.47+0.03 may indicate an expansion motion of gas.
  
We conducted the spectral analyses of these molecular lines (Figures \ref{fig:HC5N_line}--\ref{fig:NCOM_line}) by the MCMC method in CASSIS.   
Since we have clearly identified only one line or a few blended lines for these molecules, we treated their column density, FWHM, and $V_{\rm {LSR}}$ as free parameters, and fixed the excitation temperature and size.
The excitation temperatures were fixed at the values derived from the CH$_{3}$CN analysis; 86.4 K (G9.62+0.19 C1), 150 K (G9.62+0.19 C2), 113 K (G9.62+0.19 C3), 194.8 K (G10.47+0.03, component 1), 137.3 K (G10.47+0.03, component 2), 100 K (G12.89+0.49), 135 K (G16.57-0.05), 205 K (G19.88-0.53 C1), and 88 K (G19.88-0.53 C2), respectively.
The size values were fixed at the sizes of the beam; 1\farcs01 (G9.62+0.19 C1), 1\farcs24 (G9.62+0.19 C2), 1\farcs93 (G9.62+0.19 C3), 0\farcs93 (G10.47+0.03), 1\farcs84 (G12.89+0.49), 1\farcs63 (G16.57-0.05), 1\farcs25 (G19.88-0.53 C1), and 1\farcs77 (G19.88-0.53 C2), respectively.
This means that the derived column densities are the beam-averaged values.
The rms noise levels were considered in the spectral fitting.
Red curves in Figures \ref{fig:HC5N_line}--\ref{fig:NCOM_line} indicate the best-fit models.
Table \ref{tab:mol_MCMC} summarizes the column density, line width, and velocity component derived by the MCMC method.
All of these lines, except for CH$_{3}$OH in G10.47+0.03, are optically thin ($\tau < 0.3$).
The optical thickness of the CH$_{3}$OH line in G10.47+0.03 is 2.09.

\begin{deluxetable*}{lccccc}
\tabletypesize{\scriptsize}
\tablecaption{Results of MCMC analysis of CH$_{3}$CN \label{tab:CH3CN_MCMC}}
\tablewidth{0pt}
\tablehead{
\colhead{Core} & \colhead{$N$ (cm$^{-2}$)} & \colhead{$T_{\rm {ex}}$ (K)} & \colhead{FWHM (km\,s$^{-1}$)} & \colhead{$V_{\rm {LSR}}$ (km\,s$^{-1}$)} & \colhead{size (\arcsec)}
}
\startdata
G9.62+0.19 C1 & ($2.8\pm0.6$)$\times10^{15}$ & $86.40\pm0.18$ & $5.04\pm0.01$ & $4.9\pm0.3$ &  $1.488\pm 0.014$ \\
G9.62+0.19 C2 & ($1.36\pm0.04$)$\times10^{16}$ & $150.02\pm0.01$ & $5.003\pm0.001$ & $3.2 \pm0.4$ & $1.497 \pm 0.005$ \\
G9.62+0.19 C3 (component 1) & ($8.304\pm0.006$)$\times10^{15}$ & $119.84\pm 0.14$ & $5.14\pm0.03$ & $5.504\pm0.004$ & $1.316\pm0.005$ \\
G9.62+0.19 C3 (component 2) & ($7.00\pm0.01$)$\times10^{16}$ & $112.6\pm0.3$ & $8.49\pm0.01$ & $5.214\pm0.008$ & $0.735\pm0.002$ \\
G10.47+0.03 (component 1)  & ($2.27\pm0.03$)$\times10^{16}$ & $194.80\pm0.03$ & $4.69\pm0.02$ & $65.792\pm0.007$ & $2.085\pm0.013$ \\ 
G10.47+0.03 (component 2)  & ($4.982\pm0.001$)$\times10^{17}$ & $137.33\pm0.03$ & $6.84\pm0.01$ & $67.09\pm0.06$ & $1.878\pm0.009$ \\
G12.89+0.49 (component 1)  & ($2.98\pm0.03$)$\times10^{16}$ & $102.84\pm0.07$ & $1.87\pm0.03$ & $37.22\pm0.05$ & $0.444\pm0.007$ \\
G12.89+0.49 (component 2)  & ($4.3\pm0.4$)$\times10^{16}$ & $100.15\pm0.19$ & $4.06\pm0.10$ & $33.145\pm0.008$ & $1.195\pm0.013$ \\
G16.57-0.05  & ($3.6\pm0.3$)$\times10^{15}$ & $134.98\pm0.05$ & $4.99\pm0.05$ & $59.45\pm0.02$ & $1.16\pm0.11$ \\
G19.88-0.53 C1& ($6.26\pm0.18$)$\times10^{15}$ & $205.020\pm0.004$ & $4.94\pm0.03$ & $45.22\pm0.02$ & $1.02\pm0.03$ \\
G19.88-0.53 C2& ($2.96\pm0.19$)$\times10^{14}$ & $88.06\pm0.02$ & $3.5\pm0.3$ & $43.94\pm0.04$ & $2.5\pm0.3$ \\
\enddata
\tablecomments{The errors indicate the standard deviation derived from the MCMC analysis.}
\end{deluxetable*}

\begin{splitdeluxetable*}{lcccccccBlccccccccBlccccccccc}
\tabletypesize{\scriptsize}
\tablecaption{Results of MCMC analysis \label{tab:mol_MCMC}}
\tablewidth{70mm}
\tablehead{
\colhead{} & \multicolumn{3}{c}{HC$_{5}$N} & \colhead{} & \multicolumn{3}{c}{CH$_{3}$OH} & \colhead{} & \colhead{} & \multicolumn{3}{c}{CH$_{3}$OCH$_{3}$} & \colhead{} & \multicolumn{3}{c}{HNCO} & \colhead{} &\colhead{} & \multicolumn{3}{c}{CH$_{2}$CHCN} & \colhead{}  & \multicolumn{3}{c}{CH$_{3}$CH$_{2}$CN} \\
\cline{2-4} \cline{6-8} \cline{11-13} \cline{15-17} \cline{20-22} \cline{24-26} 
\colhead{} & \colhead{$N$} & \colhead{FWHM} & \colhead{$V_{\rm {LSR}}$} & \colhead{} & \colhead{$N$} & \colhead{FWHM} & \colhead{$V_{\rm {LSR}}$} & \colhead{} & \colhead{} & \colhead{$N$} & \colhead{FWHM} & \colhead{$V_{\rm {LSR}}$} & \colhead{} & \colhead{$N$} & \colhead{FWHM} & \colhead{$V_{\rm {LSR}}$} & \colhead{} &\colhead{} & \colhead{$N$} & \colhead{FWHM} & \colhead{$V_{\rm {LSR}}$} & \colhead{} & \colhead{$N$} & \colhead{FWHM} & \colhead{$V_{\rm {LSR}}$} \\
\colhead{Core} & \colhead{(cm$^{-2}$)} & \colhead{(km\,s$^{-1}$)} & \colhead{(km\,s$^{-1}$)} & \colhead{} & \colhead{(cm$^{-2}$)} & \colhead{(km\,s$^{-1}$)} & \colhead{(km\,s$^{-1}$)} & \colhead{} & \colhead{Core} & \colhead{(cm$^{-2}$)} & \colhead{(km\,s$^{-1}$)} & \colhead{(km\,s$^{-1}$)} & \colhead{} & \colhead{(cm$^{-2}$)} & \colhead{(km\,s$^{-1}$)} & \colhead{(km\,s$^{-1}$)} & \colhead{} & \colhead{Core} & \colhead{(cm$^{-2}$)} & \colhead{(km\,s$^{-1}$)} & \colhead{(km\,s$^{-1}$)} & \colhead{} & \colhead{(cm$^{-2}$)} & \colhead{(km\,s$^{-1}$)} & \colhead{(km\,s$^{-1}$)} 
}
\startdata
G9.62+0.19 C1 & ... & ... & ... & & ($1.01\pm0.05$)$\times10^{18}$ & $2.56\pm0.17$ & $5.13\pm0.06$ & & G9.62+0.19 C1 & ($3.140\pm0.006$)$\times10^{16}$ & $2.51\pm0.01$ & $4.507\pm0.006$ & & ... & ...&...& & G9.62+0.19 C1 & ... & ... & ... & & ... & ... & ...  \\
G9.62+0.19 C2 & ($5.37\pm0.07$)$\times10^{14}$ & $3.38\pm0.04$ & $3.49\pm0.02$ & & ($2.33\pm0.02$)$\times10^{18}$ & $4.50\pm0.05$ & $1.49\pm0.02$ & & G9.62+0.19 C2 & ($2.216\pm0.002$)$\times10^{17}$ & $4.57\pm0.01$ & $2.228\pm0.003$ & & ($2.367\pm0.001$)$\times10^{16}$ & $6.777\pm0.004$ & $1.778\pm0.002$ & & G9.62+0.19 C2& ($2.20\pm0.02$)$\times10^{15}$ & $2.37\pm0.04$ & $2.69\pm0.02$ & & ($6.81\pm0.02$)$\times10^{15}$ & $4.03\pm0.01$ & $2.042\pm0.005$ \\
G9.62+0.19 C3 & ($1.27\pm0.03$)$\times10^{14}$ & $5.31\pm0.11$ & $4.87\pm0.07$ & & ($5.4\pm0.1$)$\times10^{17}$ & $5.10\pm0.12$ & $5.49\pm0.05$ & & G9.62+0.19 C3 & ($6.90\pm0.01$)$\times10^{16}$ & $6.42\pm0.01$ & $4.775\pm0.006$ & & ($9.885\pm0.007$)$\times10^{15}$ & $7.102\pm0.006$ & $5.545\pm0.002$ & & G9.62+0.19 C3 & ($3.47\pm0.01$)$\times10^{15}$ & $5.84\pm0.05$ & $5.82\pm0.02$ & & ($1.189\pm0.001$)$\times10^{16}$ & $5.083\pm0.006$ & $5.678\pm0.002$ \\
G10.47+0.03 (component 1)  & ($4.82\pm0.02$)$\times10^{15}$\tablenotemark{a} & $11.997\pm0.002$ & $66.1000\pm0.0003$ & & ($5.517\pm0.007$)$\times10^{19}$ & $5.09\pm0.02$ & $65.97\pm0.01$ & & G10.47+0.03 (component 1) & ($1.263\pm0.004$)$\times10^{18}$ & $6.76\pm0.05$ & $63.10\pm0.09$ & & ($4.7\pm0.6$)$\times10^{17}$ & $8.3\pm0.3$ & $66.2\pm0.1$ & & G10.47+0.03 (component 1) & ($5.63\pm0.12$)$\times10^{17}$ & $9.02\pm0.09$ & $66.22\pm0.01$ & & ($6.01\pm0.17$)$\times10^{16}$ & $7.6\pm0.5$ & $65.9\pm0.1$ \\
G10.47+0.03 (component 2)  & \multicolumn{3}{c}{Fitted by single component} & & ($6.2\pm0.5$)$\times10^{20}$ & $6.4\pm0.2$ & $66.5\pm0.1$ & & G10.47+0.03 (component 2) & ($7.9\pm0.1$)$\times10^{17}$ & $8.14\pm0.13$ & $72.13\pm0.06$ & & ($6.8\pm1.1$)$\times10^{17}$ & $11.6\pm0.5$ & $66.0\pm0.2$ & & G10.47+0.03 (component 2)  & ($3.3\pm0.2$)$\times10^{17}$ & $5.05\pm0.07$ & $67.42\pm0.03$ & & ($2.4\pm0.2$)$\times10^{17}$ & $8.1\pm0.4$ & $67.1\pm0.2$  \\
G12.89+0.49 & ($1.01\pm0.05$)$\times10^{14}$ & $5.23\pm0.15$ & $32.88\pm0.13$ & & ($4.53\pm0.02$)$\times10^{18}$ & $4.07\pm0.02$ & $32.73\pm0.01$ & & G12.89+0.49 & ($2.521\pm0.001$)$\times10^{17}$ & $4.916\pm0.004$ & $33.004\pm0.001$ & & ($1.8863\pm0.0007$)$\times10^{16}$ & $7.678\pm0.003$ & $32.286\pm0.001$ & & G12.89+0.49 & ($2.45\pm0.03$)$\times10^{15}$ & $8.50\pm0.17$ & $33.14\pm0.06$ & & ($8.029\pm0.009$)$\times10^{15}$ & $5.130\pm0.007$ & $33.683\pm0.003$ \\
G16.57-0.05 & ...& ... & ... & &  ($2.86\pm0.03$)$\times10^{17}$ & $4.65\pm0.05$ & $59.36\pm0.02$ & & G16.57-0.05 & ($4.51\pm0.01$)$\times10^{16}$ & $6.51\pm0.03$ & $60.96\pm0.01$ & & ($1.524\pm0.007$)$\times10^{15}$ & $4.32\pm0.03$ & $59.69\pm0.01$ & & G16.57-0.05 &...&...&...& & ...&...&... \\
G19.88-0.53 C1 & ...&...&...& & ($8.95\pm0.03$)$\times10^{17}$ & $4.53\pm0.02$ & $46.10\pm0.01$ & & G19.88-0.53 C1 & ($4.67\pm0.02$)$\times10^{16}$ & $5.43\pm0.06$ & $45.17\pm0.02$ & & ($1.123\pm0.002$)$\times10^{16}$ & $6.97\pm0.01$ & $44.486\pm0.006$ & &G19.88-0.53 C1 &...&...&...& & ...&...&...  \\
G19.88-0.53 C2 & ...&...&...& & ...&...&...& & G19.88-0.53 C2& ($9.36\pm0.04$)$\times10^{15}$ & $3.59\pm0.05$ & $43.76\pm0.02$ & & ($6.82\pm0.04$)$\times10^{14}$ & $5.36\pm0.04$ & $43.63\pm0.01$ & & G19.88-0.53 C2 &...&...&...&  &...&...&... \\
\enddata
\tablecomments{The errors indicate the standard deviation derived from the MCMC analysis.}
\tablecomments{``...'' means non-detection.}
\tablenotetext{a}{The excitation temperature was fixed at 137 K.}
\end{splitdeluxetable*}

\section{Discussions} \label{sec:dis}

In the MCMC analyses (Section \ref{sec:spectralana}), we took the rms noise levels of the spectra into consideration, but uncertainties brought by other factors, such as the absolute calibration uncertainties, are not included. 
Typical absolute calibration uncertainty is 5\% at Band 3. 
Besides, there are other possible uncertainties such as molecular emission size and beam size. 
Taking these factors into consideration, we added 10\% errors to the errors of each column density derived by the MCMC methods in the following sections.

\subsection{Comparisons of spatial distributions and line features} \label{sec:d1}

\begin{figure*}[!th]
 \begin{center}
  \includegraphics[bb =7 95 575 780, scale=0.85]{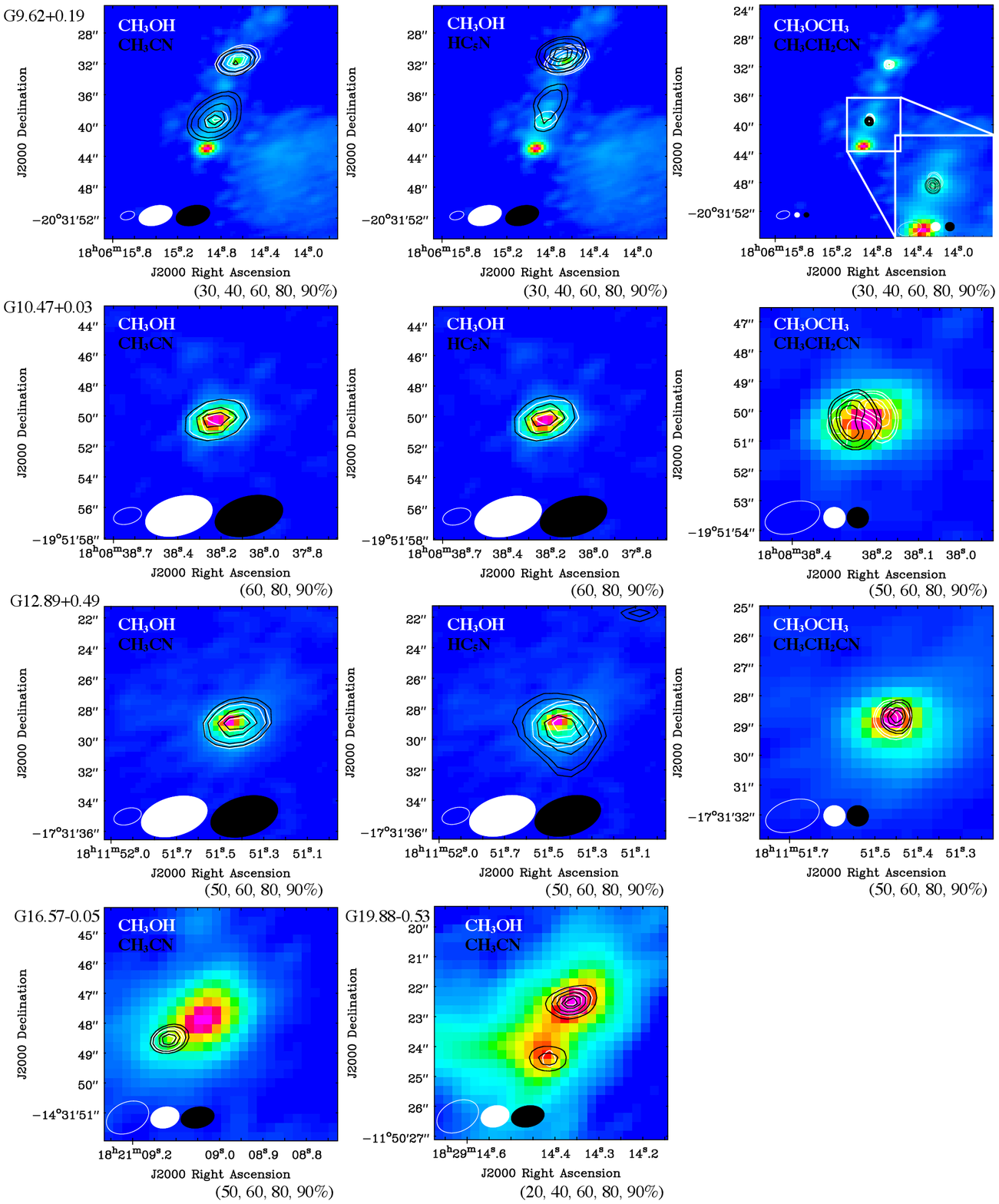}
 \end{center}
\caption{Comparison of spatial distributions among each species. The color scale shows the continuum image as the same ones of Figure \ref{fig:cont}. The white and black contours show moment 0 maps of each species denoted at the top-left corners in each panel. The contour levels are relative values of the peak intensities, and the contour levels are indicated below each panel. Three panels of the first column, the second column, and the third column show comparisons in G9.62+0.19, G10.47+0.03, and G12.89+0.49, respectively. The ellipses indicate the angular resolution; open one corresponds to the continuum images, and white and black ones correspond to the moment 0 maps of each molecule. The colors are the same ones as the contours.  \label{fig:spatial_com}}
\end{figure*}

\begin{figure}[!th]
 \begin{center}
  \includegraphics[bb =0 20 356 156, scale=0.65]{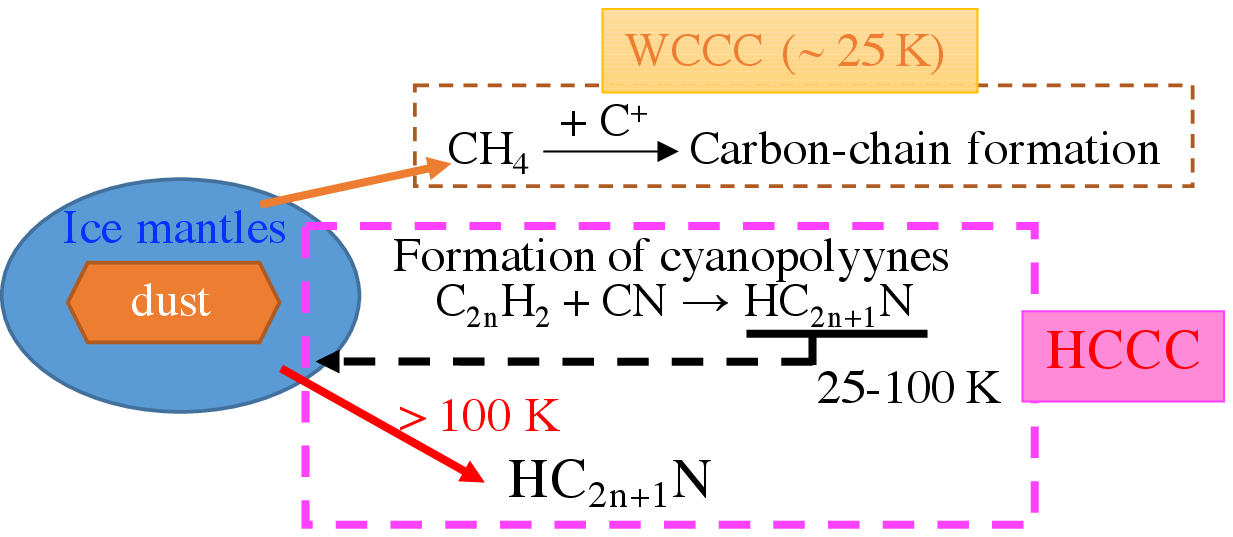}
  \end{center}
\caption{Warm Carbon-Chain Chemistry (WCCC) and Hot Carbon-Chain Chemistry (HCCC). The HCCC mechanism was based on the chemical simulations by \citet{2019ApJ...881...57T}. \label{fig:HCCC}}
\end{figure}


We compare spatial distributions among O-bearing COMs, N-bearing COMs, and HC$_{5}$N at each core. 
In these comparisons, as we have already mentioned in Section \ref{sec:mom0}, we selected molecular pairs which were observed simultaneously; CH$_{3}$OH vs. CH$_{3}$CN, CH$_{3}$OH vs. HC$_{5}$N, and CH$_{3}$OCH$_{3}$ vs. CH$_{3}$CH$_{2}$CN.
Each pair has the same angular resolutions.
In the case of G16.57-0.05 and G19.88-0.53, we only show comparisons between CH$_{3}$OH and CH$_{3}$CN, because the other molecules were not detected.

Figure \ref{fig:spatial_com} shows comparisons of spatial distributions between each pair toward the five sources.
In G9.62+0.19, the peak positions of CH$_{3}$OH and CH$_{3}$CN are well coincident with each other.
On the other hand, the peak positions between CH$_{3}$OH and HC$_{5}$N, and CH$_{3}$OCH$_{3}$ and CH$_{3}$CH$_{2}$CN may be slightly shifted, but the differences are within the beam sizes.
There is no contours of CH$_{3}$CH$_{2}$CN at the C2 position, because the relative intensity of its peak at C2 is much weaker than that at C3.

The peak positions of CH$_{3}$OH, CH$_{3}$CN, and HC$_{5}$N are well consistent with each other in G10.47+0.03 and G12.89+0.49, and their peaks are associated with the continuum cores.
The peaks of CH$_{3}$OCH$_{3}$ and CH$_{3}$CH$_{2}$CN are located at the west and east, respectively in G10.47+0.03, whereas their peaks are well consistent with the continuum peak in G12.89+0.49.

Here we simulated the line intensity of the CH$_{3}$OH line in the CASSIS software, using the results of our chemical simulations.
Details about our chemical simulations are described in Section \ref{sec:d3}.
We assume that the CH$_{3}$OH emission does not come from the hot core regions, thermal sublimation does not occur and the excitation temperature should be below 100 K.
In our chemical simulations, the maximum abundance of CH$_{3}$OH with respect to total hydrogen nuclei before its thermal sublimation is around $5\times10^{-9}$.
We further assume that the H$_{2}$ column density is $10^{24}$ cm$^{-2}$ and the CH$_{3}$OH excitation temperature is 90 K, which is the corresponding temperature at the maximum CH$_{3}$OH abundance.
With these conditions, the observed CH$_{3}$OH line is not excited and should not have been detected.
Therefore, the observed CH$_{3}$OH line should trace the hot core regions with temperatures above $\sim 100$ K in which thermal sublimation of CH$_{3}$OH occurs.

Both the CH$_{3}$OH and HC$_{5}$N lines are optically thin, except for CH$_{3}$OH in G10.47+0.03.
In addition, the CH$_{3}$OH and HC$_{5}$N emissions show single peaks toward the source positions and the emissions are almost unresolved, where the beam sizes of both observations are almost identical.
Thus, both the CH$_{3}$OH and HC$_{5}$N emission trace the hot and dense gas regions associated with the HMPOs.

These results support previous results and implications from chemical simulations \citep{2019ApJ...881...57T} and single-dish-telescope observations \citep{2021ApJ...908..100T}; 
cyanopolyynes around MYSOs exist in higher temperature regions ($\geq 85$ K) than regions where WCCC occurs ($\sim25-30$ K).
This means that cyanopolyynes can survive in hot regions where radical type carbon-chain species (e.g., CCH and CCS) are efficiently destroyed by the gas-phase reactions with O and H$_{2}$ \citep{2019ApJ...881...57T}.
Moment 0 maps (Figures \ref{fig:mom0_G9}--\ref{fig:mom0_G12}) support this picture.
The CCH emission is systematically different from those of CH$_{3}$OH and HC$_{5}$N; rather extended, clumpy features, and deficient at the continuum peaks.
In other words, the carbon-chain chemistry around MYSOs is not WCCC occurring around low-mass YSOs, but another mechanism in hot-core regions ($>100$ K).
We call this carbon-chain chemistry occurring in hot-core regions around MYSOs {\it Hot Carbon-Chain Chemistry (HCCC)}.

Figure \ref{fig:HCCC} shows a schematic view of WCCC and HCCC to distinguish them clearly.
The WCCC mechanism indicates the formation processes of carbon-chain species, starting from CH$_{4}$ at temperatures around 25 K, corresponding to the sublimation temperature of CH$_{4}$.
The HCCC mechanism represents not only gas-phase formation of cyanopolyynes around 25 K, but also successive phenomena.
In the HCCC mechanism, cyanopolyynes are formed in the warm gas by neutral-neutral reactions between C$_{2n}$H$_{2}$ and CN, adsorbed onto dust grains, and accumulated in ice mantles during the warm-up stage \citep[$25 \leq T \leq 100$ K;][]{2019ApJ...881...57T}. 
After the temperature reaches 100 K, cyanopolyynes sublimate into the gas phase and show peak abundances, whereas the radical-type carbon-chain species (e.g., CCH) are depleted due to efficient destruction in the gas phase \citep{2019ApJ...881...57T}.
We further discuss comparisons with our latest chemical simulations in Section \ref{sec:d3}.

In G16.57-0.05, the peak positions of CH$_{3}$OH and CH$_{3}$CN are well consistent with each other, but they are not coincident with the continuum core.
Since the continuum flux in G16.57-0.05 is not higher than the other sources, the spatial difference between the continuum peak and the molecular emission appears to be real.
In G19.88-0.53, the peak positions of CH$_{3}$OH and CH$_{3}$CN are well coincident each other, and they are associated with the continuum cores.

In summary, we cannot recognize clear chemical differentiation around the five MYSOs with the current angular resolutions.
In G10.47+0.03, the spatial distributions of CH$_{3}$OCH$_{3}$ and CH$_{3}$CH$_{2}$CN show different features.
As mentioned in Section \ref{sec:spectralana}, the CH$_{3}$OCH$_{3}$ line in this source likely traces the expanding shell or an accretion disk.
However, the CH$_{3}$CH$_{2}$CN line does not show such a double-peaked feature  (Figure \ref{fig:NCOM_line}).
Thus, their spatial distributions probably differ.
Future higher angular-resolution data are necessary to confirm the possible chemical differentiation.

\subsection{Comparison of molecular composition} \label{sec:d2}

\begin{figure*}[!th]
 \begin{center}
  \includegraphics[bb=0 30 700 500, scale=0.7]{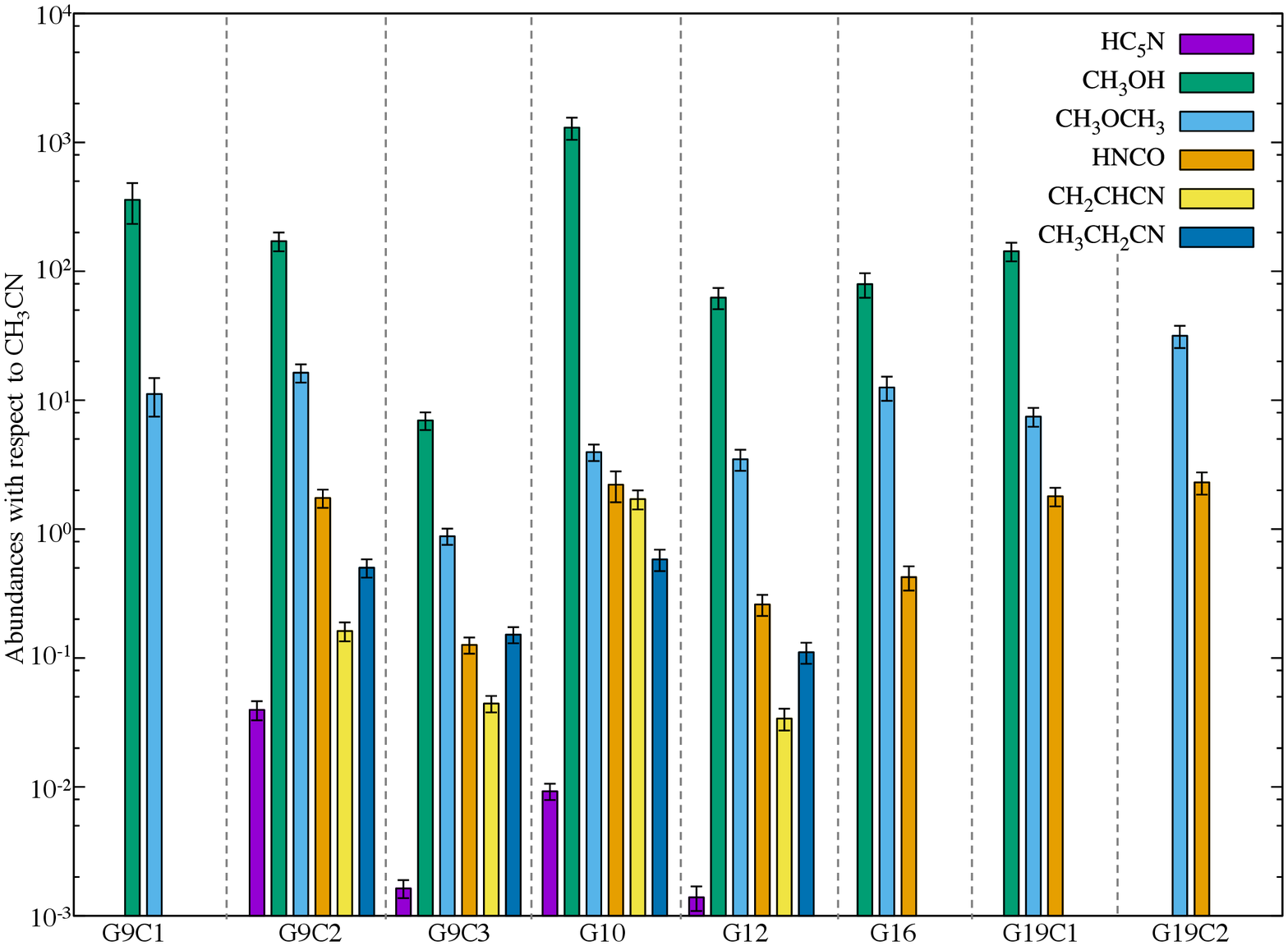}
 \end{center}
\caption{Comparison of chemical compositions with respect to CH$_{3}$CN. \label{fig:chemical_com}}
\end{figure*}

The emission of HC$_{5}$N and COMs show point-like distributions.
Although these datasets are inhomogeneous, it is still meaningful to compare the chemical composition among the different sources using the observed molecules.
We derive molecular abundances with respect to CH$_{3}$CN (=$N$($X$)/$N$(CH$_{3}$CN), where $X$ denotes molecular species) and compare them among each core.
We used CH$_{3}$CN as a reference, because we consider that the spectral analysis of this species is most reliable due to its $K-$ladder lines (Section \ref{sec:spectralana}).
We added Components 1 and 2, when we applied two-component fitting in the spectral analyses (Section \ref{sec:spectralana}).

Figure \ref{fig:chemical_com} shows comparisons of chemical composition among each core.
This figure suggests chemical diversity among MYSOs.
The cores where we detected the HC$_{5}$N line are associated with all of the molecular species.
For almost all of the species, the abundances with respect to CH$_{3}$CN vary by more than one order of magnitude among the cores, while the CH$_{3}$CH$_{2}$CN abundances show differences within a factor of $\sim5$.
The CH$_{2}$CHCN abundance with respect to CH$_{3}$CN in G10.47+0.03 is higher than that of CH$_{3}$CH$_{2}$CN, which is an opposite trend in the other cores.

We investigate physical properties of sources between HC$_{5}$N-detected (G9.62+0.19, G10.47+0.03, and G12.89+0.49) and HC$_{5}$N-undetected (G16.57-0.05 and G19.88-0.53) sources.
The physical properties derived by \citet{2018MNRAS.473.1059U}, summarized in Table \ref{tab:source} of this paper, show similar values of $T_{\rm {dust}}$ and $N$(H$_{2}$), which are most related to chemistry, among all of the five sources.
However, \citet{2018MNRAS.473.1059U} derived these parameters for the same beam size, and then the linear scales were different among each source due to different source distances.
We then adopted modified blackbody models for a linear scale of 0.5 pc toward all of the sources, as explained in Appendix \ref{sec:a3}.
Figure \ref{fig:sed} in Appendix \ref{sec:a3} shows plots of infrared fluxes and the modified blackbody models which can reproduce the observed fluxes best, and Table \ref{tab:sed} summarizes the fitting parameters.

The cold-component temperatures ($T_{c}$) are almost consistent within their errors among all of the sources except for G10.47+0.03 (Table \ref{tab:sed}).
The luminosity ($L$) in G12.89+0.49 associated with the HC$_{5}$N emission is consistent with that in G16.57-0.05, where HC$_{5}$N has not been detected, within their errors.
Thus, we cannot confirm an importance of luminosity for detection of HC$_{5}$N.

However, the different HC$_{5}$N abundances among the four cores (G9.62+0.19 C2, G9.62+0.19 C3, G10.47+0.03, and G12.89+0.49) by more than one order of magnitude may be related to the luminosity; G12.89+0.49 has the lowest HC$_{5}$N abundances and the lowest luminosity among the three sources.
In contrast, the HC$_{5}$N abundance at G9.62+0.19 C2 is highest among the four cores.
This core was proposed to be a late HMC or an HC \ion{H}{2} region \citep{2017ApJ...849...25L}.
If this source is an HC \ion{H}{2} region, the UV radiation field should be stronger compared with the other cores.
This implication may be consistent with the suggestion by \citet{2017A&A...605A..57F}; a high cosmic ray ionization rate ($\zeta=4\times10^{-14}$ s$^{-1}$) could explain their observational results of the higher HC$_{5}$N abundance with respect to HC$_{3}$N in the protocluster OMC-2 FIR\,4 and FIR\,5.
The inner density structure should be related to penetration of energetic particles and photons, and we need higher angular resolution observations to confirm their effects.
The non-detection of HC$_{5}$N at G9.62+0.19 C1, corresponding to an UC  \ion{H}{2} region \citep{2017ApJ...849...25L}, means that HC$_{5}$N may be efficiently destroyed at this evolutionary stage.

Although we cannot recognize significant differences in luminosity between the HC$_{5}$N-detected and HC$_{5}$N-undetected sources, the ratio of solid angles between warm and cold components ($\Omega_{w}/\Omega_{c}$) are different between them more clearly.
We found that the HC$_{5}$N-detected sources have larger $\Omega_{w}/\Omega_{c}$ ratios compared to the other group.
Moreover, temperatures of cold components are relatively higher in the HC$_{5}$N-detected sources.
These results imply that the hot regions are likely more extended around MYSOs where HC$_{5}$N has been detected.
This can be interpreted as that HC$_{5}$N and N-bearing COMs are detected from more evolved MYSOs.
In summary, hot region is likely important for detection of HC$_{5}$N around MYSOs, namely the HCCC mechanism (Figure \ref{fig:HCCC}), and moderate high dose of energetic photons or particles (e.g., UV radiation, cosmic rays) seems to further enhance the HC$_{5}$N abundance.

\subsection{Comparison between observational results and simulations} \label{sec:d3}

\begin{figure*}[!th]
 \begin{center}
  \includegraphics[bb= -50 200 666 684, width=\textwidth]{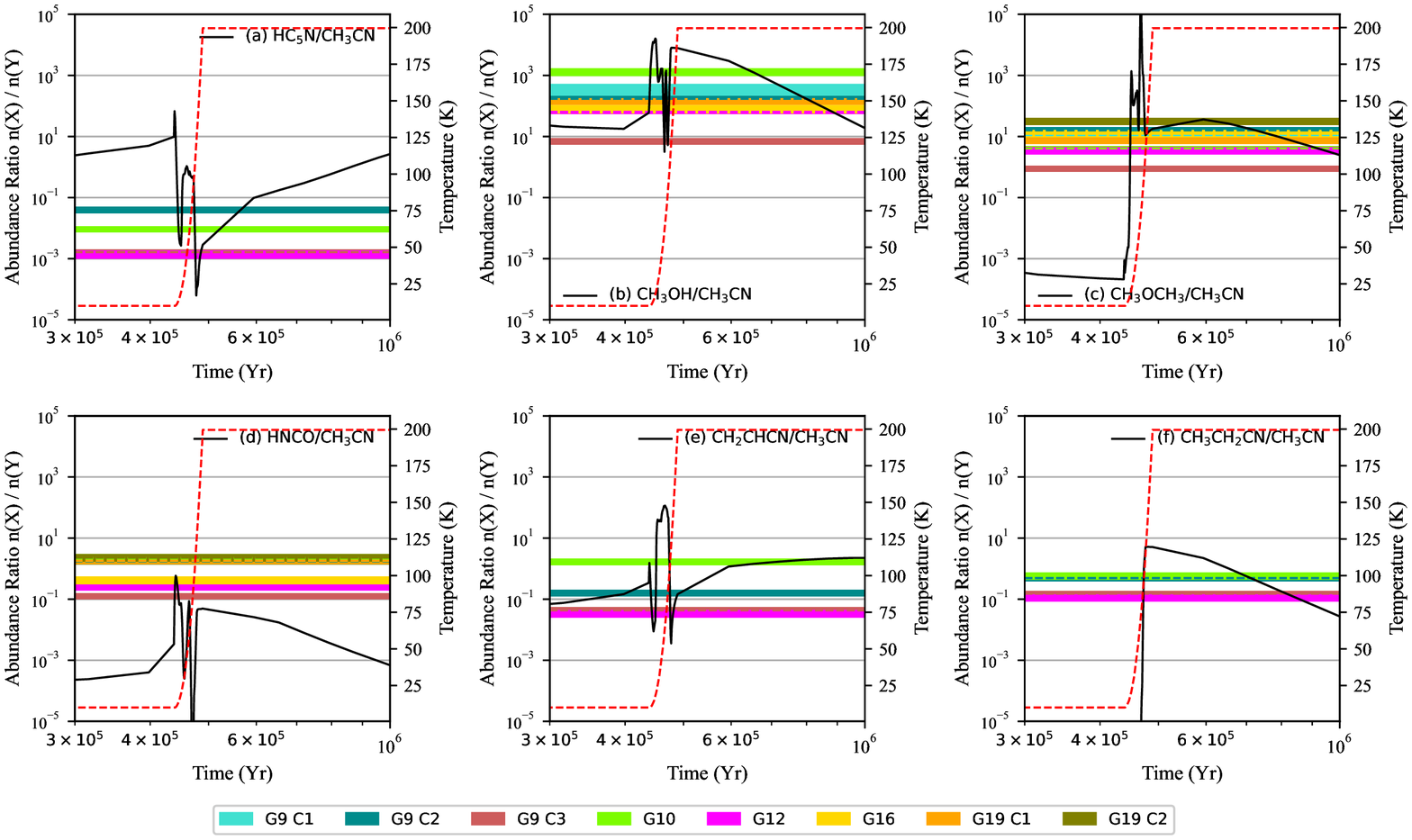}
 \end{center}
\caption{Comparison between the observed molecular abundances with respect to CH$_{3}$CN at each core (horizontal color lines) and the modeled ones (black solid lines). The red dashed lines indicate the temperature evolution. \label{fig:obs_model}}
\end{figure*}

In this subsection, we will compare results of the observations and our chemical simulations. 
To simulate the abundances of the detected carbon-chain species, O-bearing and N-bearing COMs in the observed MYSOs, we have used the state-of-the-art three-phase chemical code Nautilus \citep{2016MNRAS.459.3756R}. 
Nautilus computes the time evolution of abundances of chemical species for a given set of physical conditions and chemical parameters. 
It simulates chemistry in the gas phase, grain surface, and grain mantle by considering various exchange processes such as adsorption of gas-phase species onto grain surfaces, the thermal and non-thermal desorption of species from grain surface into the gas-phase, and the surface-mantle and mantle-surface exchange of species. 
Our gas-phase chemical network is based on the public chemical network kida.uva.2014 \citep{2015ApJS..217...20W} with updates on chemistry of O-bearing and N-bearing COMs \citep{2018MNRAS.473L..59M, 2018ApJS..237....3S}. 
The surface chemical network is based on the one from \citet{2007A&A...467.1103G} with several additional processes and reactions \citep{2015MNRAS.447.4004R, 2018MNRAS.473L..59M, 2018ApJS..237....3S}.

To simulate the physical conditions in MYSOs, we have used the two-stage fast-warm-up model: free-fall collapse, followed by a dynamically static warm-up stage \citep{2006A&A...457..927G}.
The evolution of the physical parameters and initial elemental abundances were set at the same values as in \citet{2019ApJ...881...57T}.
In the free-fall collapse stage, the temperature is kept at 10 K, and the initial gas density is $10^{4}$ cm$^{-3}$.
The gas density increases to $10^{7}$ cm$^{-3}$ during the free-fall collapse, which lasts for $\sim 4.5\times10^{5}$ yr.
According to the density increase, the visual extinction ($A_{\rm {v}}$) increases from 5 mag to 500 mag.
After the density reaches $10^{7}$ cm$^{-3}$, the warm-up stage starts and the temperature rises from 10 K to 200 K.  
The heating timescale is $5\times10^{4}$ yr, which represents a typical evolutionary timescale of formation of high-mass stars \citep{2006A&A...457..927G}.
After the temperature reaches 200 K at $\sim 5.9\times10^{5}$ yr, the temperature is kept at 200 K until $10^{7}$ yr (the hot-core stage).
The temperature evolution is indicated as red dashed lines in Figure \ref{fig:obs_model}.
The dust grain size is fixed at 0.1 \micron.
The cosmic ray ionization rate ($\zeta$) of $1.3\times10^{-17}$ s$^{-1}$ was adopted.
We also run simulations with high cosmic-ray ionization rates ($3.0\times10^{-16}$ s$^{-1}$ and $4.0\times10^{-14}$ s$^{-1}$), as previous studies did \citep{2017A&A...605A..57F, 2019ApJ...881...57T}.
We find that models with higher cosmic-ray ionization rates cannot reproduce the observational results reasonably.
Figure \ref{fig:modelcomp} in Appendix \ref{sec:a3} shows comparisons of chemical simulations with three different cosmic-ray ionization rates.
Here, we exclude these two results, and show results with $\zeta=1.3\times10^{-17}$ s$^{-1}$.

Figure \ref{fig:obs_model} shows comparisons of abundances between the observations (horizontal color lines) and the chemical simulations (black lines).
Since our target sources include MYSOs, we focus on the warm-up and hot-core stages.

Panel (a) of Figure \ref{fig:obs_model} shows the time evolution of the HC$_{5}$N/CH$_{3}$CN abundance ratio.
The observed HC$_{5}$N/CH$_{3}$CN abundance ratios agree with the simulation around ($4.45-4.51$)$\times10^5$ yr and ($4.8-6.0$)$\times10^5$ yr, and corresponding temperature ranges are $\sim 12-19$ K and $\sim 160-200$ K, respectively.
Here, we estimated the upper limit of line intensity of the observed HC$_{5}$N line under low temperature ($T<20$ K) conditions.
In the estimation, the HC$_{5}$N column density is set at $1\times10^{14}$ cm$^{-2}$.
This column density was derived from the highest HC$_{5}$N abundance before the temperature reaches 20 K including free-fall collapse stage ($\simeq10^{-10}$ with respect to the number density of total hydrogen nuclei) and the assumed H$_{2}$ column density of $10^{24}$ cm$^{-2}$, which is a typical value in these high-mass star-forming regions (see Table \ref{tab:source}).
Other assumptions are as follows; $T_{\rm {ex}}=12-19$ K, FWHM = 3 km s$^{-1}$, and no beam dilution.
With these parameters, the observed HC$_{5}$N ($J=35-34$) line is not excited, and the upper limit of intensity of this HC$_{5}$N line is found to be $\sim55$ mK.
This means that the HC$_{5}$N line should not have been detected in these data sets with noise levels around 0.2 K (Figure \ref{fig:HC5N_line}), if HC$_{5}$N exists in such low temperature regions.
Hence, the age of ($4.45-4.51$)$\times10^5$ yr is not suitable for our target sources, and the most reasonable agreement between the observations and the simulation is ($4.8-6.0$)$\times10^5$ yr ($T \approx 160-200$ K).
We also conducted the same test for the other molecular lines described above, and constrained the ages in which the modeled abundance ratios agree with the observed ones.
Low-temperature conditions ($<100$ K) are rejected for all of the molecules.
In the following parts, we mention only the best chemical ages which are constrained by these tests.

At ($4.8-6.0$)$\times10^5$ yr ($T \approx 160-200$ K), both CH$_{3}$CN and HC$_{5}$N sublimate from dust grains into the gas phase by the thermal desorption mechanism.
The HC$_{5}$N molecules are formed in the gas phase, adsorbed onto dust grains, and accumulated in ice mantles during the starless-core and warm-up stages before the temperature reaches its sublimation temperature ($\sim100$ K).
Hence, we conclude that HC$_{5}$N detected around the MYSOs comes from dust grains, and exists in the hot core regions ($T>100$ K). 
These results confirm that cyanopolyyne chemistry around MYSOs cannot be explained by WCCC, and we need HCCC (c.f., Figure \ref{fig:HCCC}).
Regarding CH$_{3}$CN, the dust-surface radical-radical reaction between NH and H$_{2}$CCN produces the gas-phase CH$_{3}$CN, which is adsorbed onto dust grains again, in the starless-core stage and warm-up stage before the temperature reaches its sublimation temperature ($\sim 100$ K).

Panel (b) of Figure \ref{fig:obs_model} shows results of the CH$_{3}$OH/CH$_{3}$CN abundance ratios.
The observed abundance ratios toward all of the sources can be reproduced by the simulation around $6.54\times10^{5}$--$1.1\times10^{6}$ yr ($T=200$ K).
The CH$_{3}$OH molecules sublimate from dust grains via the thermal desorption mechanism, namely hot core chemistry.
These CH$_{3}$OH molecules coming from dust grains are formed in the starless-core stage; the CH$_{3}$OH molecules are formed by the dust-surface reaction between CH$_{3}$NH$_{2}$ and CH$_{3}$O to produce CH$_{3}$OH and CH$_{2}$NH$_{2}$. 
Some fractions of the formed two molecules desorb into the gas phase, and others remain on dust surfaces.

Panel (c) of Figure \ref{fig:obs_model} shows the CH$_{3}$OCH$_{3}$ case.
The observations and simulation agree around $6.5\times10^{5}$--$1.16\times10^{6}$ yr ($T=200$ K).
During these ages, the thermal sublimation of CH$_{3}$OCH$_{3}$ is the most dominant production pathway.
In the starless-core stage, this molecule is formed in ice mantle by the association reaction between CH$_{3}$ and CH$_{3}$O, and the formed CH$_{3}$OCH$_{3}$ molecules move to dust surfaces or gas phase so that the equilibrium state realizes.
The two O-bearing COMs can be reproduced around similar ages  ($6.5\times10^{5}$--$1.1\times10^{6}$ yr) with a temperature of 200 K, implying that both of the species are produced by the hot core chemistry.

Panel (d) shows the results of HNCO. 
We found that the observed HNCO/CH$_{3}$CN abundance ratios cannot be reproduced during the warm-up and hot-core stages, but marginally agree with the low-temperature regime ($T\approx12$ K), which is unlikely for our target sources.
This may mean that HNCO is significantly affected by the shock chemistry, which we did not take into consideration in the chemical simulation.
In fact, \citet{2019ApJ...881...32B} showed that the HNCO abundance is strongly enhanced during the sputtering and redeposition regimes.
We have already mentioned the possibility that the HNCO emission comes from outflows in Section \ref{sec:d1}.

Panels (e) and (f) of Figure \ref{fig:obs_model} show the results of CH$_{2}$CHCN and CH$_{3}$CH$_{2}$CN, respectively.
The observed CH$_{2}$CHCN/CH$_{3}$CN abundance ratios can be reproduced around ($4.8-7.2$)$\times10^{5}$ yr, when temperatures are $\sim140-200$ K, and the CH$_{3}$CH$_{2}$CN/CH$_{3}$CN abundance ratios agree with the simulation around (7--8)$\times10^{5}$ yr ($T=200$ K).
Regarding CH$_{2}$CHCN, its thermal sublimation is responsible for reproduction of the observed abundances.
CH$_{2}$CHCN is formed in the gas phase in the starless-core stage by the reaction of ``CN + C$_{2}$H$_{4}$ $\rightarrow$ CH$_{2}$CHCN + H'' and the electron recombination reactions of C$_{2}$H$_{4}$CN$^{+}$ and C$_{3}$H$_{3}$NH$^{+}$.
The formed CH$_{2}$CHCN molecules are adsorbed onto dust grains and accumulated in ice mantles.

Formation of CH$_{3}$CH$_{2}$CN is not efficient in the starless-core stage, and it can be efficiently formed in the warm-up stage. 
The CH$_{3}$CH$_{2}$CN abundance steeply increases at $\sim 4.62\times10^{5}$ yr ($T\approx47$ K).
At that time, this molecule is efficiently formed by ``JH + JCH$_{3}$CHCN  $\rightarrow$ CH$_{3}$CH$_{2}$CN'', where ``J'' denotes species on dust surface.
During the age when the observations and simulation agree ($\sim7\times10^{5}$ yr), the most dominant formation pathway in our simulation is the following gas-phase reaction;
\begin{equation} \label{equ:reac_CH3CH2CN}
{\rm HCO}+{\rm CH}_{3}{\rm CHCN} \rightarrow {\rm CH}_{3}{\rm CH}_{2}{\rm CN} + {\rm CO}.
\end{equation}
This reaction becomes dominant from $\sim4.88\times10^{5}$ yr, when the temperature reaches around 190 K, to $1.05\times10^{6}$ yr ($T=200$ K).
In summary, the observed abundances of the N-bearing COMs best agree with the simulation around $7\times10^{5}$ yr.

We found that the observed abundances of all of the molecules, except for HNCO, toward the five MYSOs can be explained by the hot-core stage when the temperature exceeds 100 K.
Although the best agreement ages are ($\sim6-7$)$\times10^{5}$ yr, the important point is not the absolute age but the corresponding temperature.
All of the observed molecular abundances, except for HNCO, agree with the model at the temperature between 160 K and 200 K.
These results support the HCCC mechanism, which is necessary for reproducing the observed HC$_{5}$N abundances around MYSOs.

\section{Conclusions} \label{sec:con}

We present ALMA Band 3 data of the carbon-chain species and COMs toward the five MYSOs, and demonstrate chemical simulations of hot-core models with a warm-up stage. 
We investigate relationships between carbon-chain species and COMs around the MYSOs with spatially-resolved data, and relationships between chemical composition and physical conditions.
Main conclusions of this paper are as follows.

\begin{enumerate}
\item The HC$_{5}$N ($J=35-34$) line has been detected from the three MYSOs, where the N-bearing COMs (CH$_{2}$CHCN and CH$_{3}$CH$_{2}$CN) have been detected.
The HC$_{5}$N line shows compact emission associated with the continuum core.
Its spatial distributions are consistent with those of COMs, which implies that the HC$_{5}$N line comes from hot-core regions with temperatures above 100 K.
On the other hand, the CCH emission is depleted at the continuum core and hot-core regions.

\item We do not find clear evidence for the chemical differentiation between O-bearing COMs and N-bearing COMs with the current angular resolutions.
In G10.47+0.03, the CH$_{3}$OCH$_{3}$ emission shows a ring-like distribution and a double-peaked feature in its spectrum, suggestive of an expanding shell or an accretion disk shock.
On the other hand, the emission of the other molecules does not show such features.

\item We derive physical parameters within 0.5 pc by the two-temperature modified blackbody model, and investigate the relationship with chemical composition. 
We did not recognize any clear difference in temperature and bolometric luminosity between the three MYSOs associated with the HC$_{5}$N emission and the other two MYSOs without it.
The solid angle ratios between warm and cold components ($\Omega_{w}/\Omega_{c}$) are higher in the MYSOs associated with HC$_{5}$N and N-bearing COMs.
These results indicate that the hot regions are much extended around the MYSOs where HC$_{5}$N and N-bearing COMs have been detected.

\item The observed molecular abundances with respect to CH$_{3}$CN around the target MYSOs agree with the chemical simulations around (6--7)$\times10^{5}$ yr, at which the temperature reaches 200 K, namely the hot-core chemistry phase.
These results support that the HC$_{5}$N emission comes from the hot-core regions with temperatures above 100 K. 
In such high temperature regions, radical-type carbon-chain species (e.g., CCH) are predicted to be efficiently destroyed in the gas phase.
This can be seen in the moment 0 maps, because the CCH emission is depleted at the peaks of the HC$_{5}$N and COMs emission, as well as the continuum emission.

\item We propose {\it Hot Carbon-Chain Chemistry (HCCC)} to explain the observed features of the carbon-chain species around MYSOs.
In HCCC, cyanopolyynes (HC$_{2n+1}$N) sublimate from dust grains with temperatures above 100 K, while the radical-type species are deficient in such hot regions.
The HCCC mechanism and hot core chemistry, which is rich with COMs, could occur simultaneously, and they are not mutually exclusive.
\end{enumerate}

\begin{acknowledgments}
This paper makes use of the following ALMA data: ADS/JAO.ALMA\#2018.1.00424.S. 
ALMA is a partnership of ESO (representing its member states), NSF (USA) and NINS (Japan), together with NRC (Canada), MOST and ASIAA (Taiwan), and KASI (Republic of Korea), in cooperation with the Republic of Chile. 
The Joint ALMA Observatory is operated by ESO, AUI/NRAO and NAOJ. 
Data analysis was in part carried out on the Multi-wavelength Data Analysis System operated by the Astronomy Data Center (ADC), National Astronomical Observatory of Japan.
Based on analysis carried out with the CASSIS software \citep[\url{http://cassis.irap.omp.eu};][]{2015sf2a.conf..313V}, and JPL and CDMS spectroscopic databases. 
CASSIS has been developed by IRAP-UPS/CNRS. 

K.T. is supported by JSPS KAKENHI grant No. JP20K14523.
K. T. appreciates NAOJ Overseas Visit Program for Young Researchers (NINS) to support travel funding for visiting Max-Planck-Institut f\"{u}r Extraterrestrische Physik. 
K.T. thanks Dr. Jorma Sakari Harju for his help on creating the modified blackbody model. 
L.M. acknowledges the financial support of DAE and DST-SERB research grants (SRG/2021/002116 and MTR/2021/000864) of the Government of India.
S.T. is supported by JSPS KAKENHI Grant Numbers JP21H00048 and JP21H04495.
ZYL is supported in part by NASA 80NSSC20K0533 and NSF AST-1910106. 
\end{acknowledgments}
%

\vspace{5mm}
\facilities{Atacama Large Millimeter/submillimeter Array (ALMA)}
\software{Common Astronomy Software Applications package \citep[CASA;][]{2007ASPC..376..127M}, CASSIS \citep{2015sf2a.conf..313V}}
%
%
\appendix

\section{Optical depth of molecular lines} \label{sec:a1}

Table \ref{tab:tau} summarizes the optical depth of lines of CH$_{3}$CN for the best-fit models derived by the MCMC method.

\begin{deluxetable*}{lcccccccc}
\tabletypesize{\footnotesize}
\tablecaption{Optical depth of the CH$_{3}$CN lines \label{tab:tau}}
\tablewidth{0pt}
\tablehead{
\colhead{Lines} & \colhead{G9.62+0.19 C1} & \colhead{G9.62+0.19 C2} & \colhead{G9.62+0.19 C3} & \colhead{G10.47+0.03} &\colhead{G12.89+0.49} & \colhead{G16.57-0.05} & \colhead{G19.88-0.53 C1} & \colhead{G19.88-0.53 C2}
}
\startdata
$K$=0 & 0.067 & 0.039 & 0.366, 2.21 & 0.300, 11.75 & 5.24, 3.14 & 0.134 & 0.067 & 0.039 \\
$K$=1 & 0.062 & 0.035 & 0.331, 1.99 & 0.282, 10.71 & 4.69, 2.81 & 0.122 & 0.062 & 0.035 \\
$K$=2 & 0.049 & 0.024 & 0.242, 1.44 & 0.221, 8.02 & 3.33, 1.98 & 0.091 & 0.049 & 0.024 \\
$K$=3\tablenotemark{a} & 0.064 & 0.024 & 0.274, 1.596 & 0.280, 5.42 & 3.58, 2.12 & 0.106 & 0.064 & 0.024 \\
\enddata
\tablecomments{In the case two values are listed, they are values for ``component 1'' and ``component 2'', respectively.}
\tablenotetext{a}{The $K$=3 transition has degeneracy ($K=-3\rightarrow3$ and $3\rightarrow-3$). The optical depths of these two transitions are the same. The listed values are the sum of these two degenerate transitions.}
\end{deluxetable*}

\section{Channel maps of CH$_{3}$OCH$_{3}$ in G10.47+0.03} \label{sec:a2}

\begin{figure*}[!th]
 \begin{center}
  \includegraphics[bb =0 15 461 455, scale=1.08]{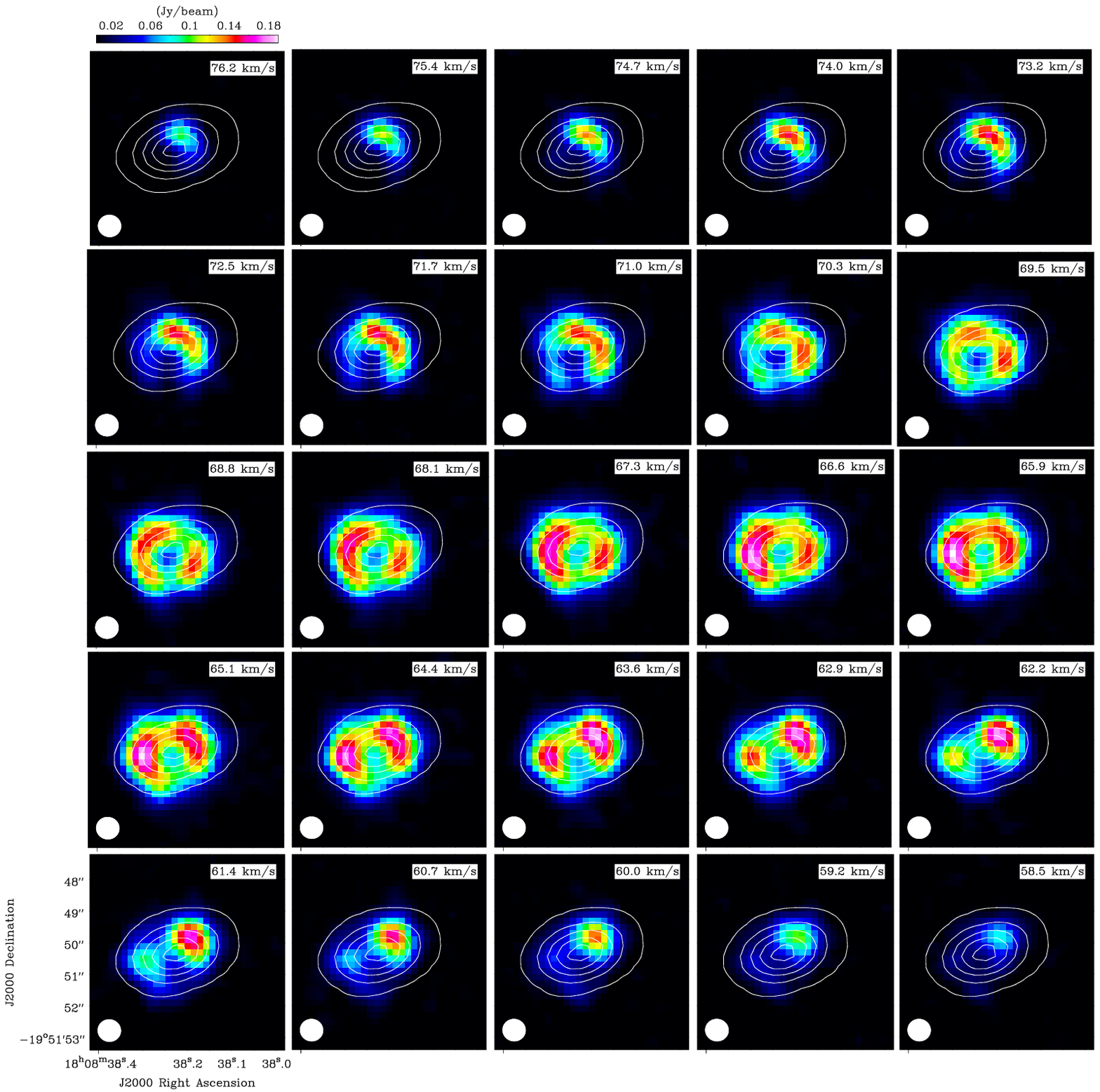}
 \end{center}
\caption{Channel maps of CH$_{3}$OCH$_{3}$ in G10.47+0.03. The white contours indicate the continuum emission (from $20\sigma$ to $180\sigma$ in $40\sigma$ step).\label{fig:G10_CH3OCH3_channel}}
\end{figure*}

Figure \ref{fig:G10_CH3OCH3_channel} shows channel maps of the CH$_{3}$OCH$_{3}$ line toward G10.47+0.03.
In the highly redshifted velocity range (76--75 km\,s$^{-1}$), the emission is located only toward the  northwest, while emission of lower-velocity components is also present towards the northeast, east and southeast.
In the highly blueshifted velocity range (59--58 km\,s$^{-1}$), the emission is again located to the northwest only.
A rotating and expanding shell motion plus skewed emission distribution toward the northwest could reproduce the observed velocity structure.
Detailed analysis using our model will be discussed in a future paper.

\section{Modified blackbody model} \label{sec:a3}

\begin{figure*}
 \begin{center}
 \includegraphics[bb =0 20 544 307, scale=0.78]{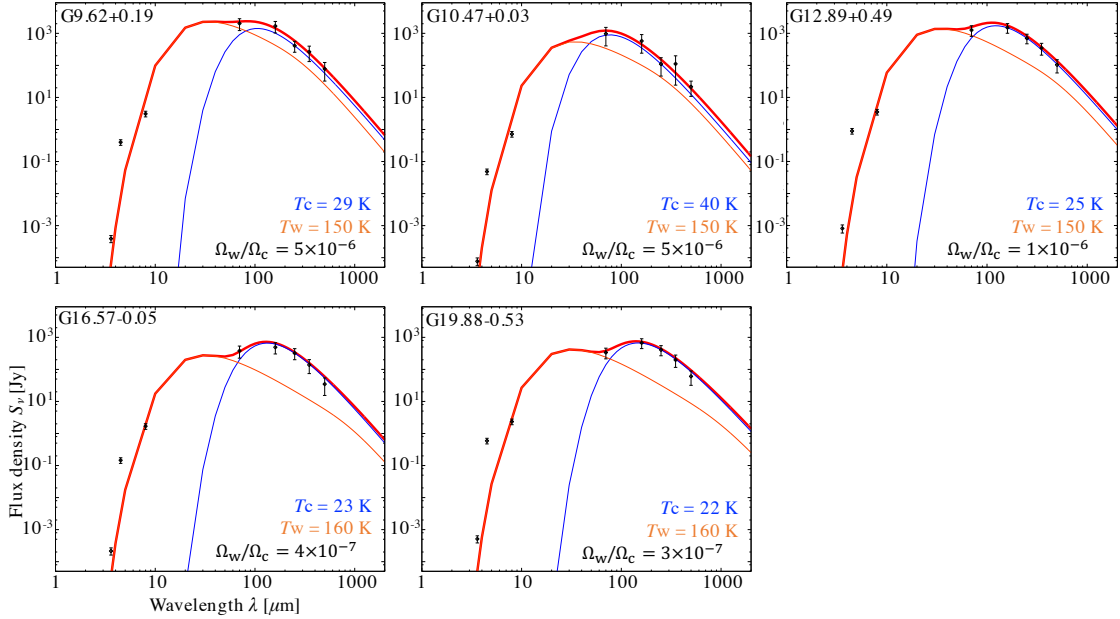}
 \end{center}
\caption{Plots of infrared fluxes and the modified blackbody models which can reproduce the observed fluxes best. Blue and orange curves indicate cold and warm components, respectively. Red curves show sums of the cold and warm components. Black points are data points taken from the archival data. $T_{c}$ and $T_{w}$ indicate temperatures of cold and warm components, respectively.  $\Omega_{w}/\Omega_{c}$ indicates applied solid angle ratios between warm and cold components. \label{fig:sed}}
\end{figure*}

\begin{deluxetable*}{lcccccccc}
\tablecaption{Flux density at each wavelength \label{tab:fluxdensity}}
\tablewidth{0pt}
\tablehead{
\colhead{Souce} & \colhead{3.6\micron} & \colhead{4.5\micron} & \colhead{8\micron} &\colhead{70\micron} & \colhead{160\micron} & \colhead{250\micron} & \colhead{350\micron} & \colhead{500\micron}
}
\startdata
G9.62+0.19   & 3.9 (0.9)$\times10^{-4}$ & 0.40 (0.08) & 3.1 (0.6) & 2037 (818) & 1688 (661) & 411 (154) & 265 (133) & 78 (45) \\
G10.47+0.03 & 7.7 (2.1)$\times10^{-5}$ & 0.049 (0.010) & 0.72 (0.14) & 970 (564) & 578 (337) & 110 (64) & 112 (87) & 22 (11)  \\
G12.89+0.49 & 8.1 (2.3)$\times10^{-4}$ & 0.89 (0.18) & 3.5 (0.7) & 1262 (460) & 1509 (504) & 704 (234) & 348 (139) & 104 (46) \\
G16.57-0.05 & 2.1 (0.5)$\times10^{-4}$  & 0.15 (0.03) & 1.7 (0.3) & 380 (155) & 491 (188) & 322 (122) & 138 (64) & 35 (19) \\
G19.88-0.53 & 5.1 (1.3)$\times10^{-4}$  & 0.59 (0.12) &2.3 (0.5) & 337 (123) & 675 (229) & 409 (143) & 198 (82) & 61 (29) \\
\enddata
\tablecomments{Unit is Jy. The values in parenthesis are uncertainties. These errors were calculated from the rms noise levels of images, and we included assumed 20\% absolute calibration errors.}
\end{deluxetable*}

\begin{deluxetable*}{lcccc}
\tablecaption{Physical parameters derived from fitting by the modified blackbody model \label{tab:sed}}
\tablewidth{0pt}
\tablehead{
\colhead{Souce} & \colhead{$T_{c}$ (K)} & \colhead{$T_{w}$ (K)} & \colhead{$\Omega_{w}/\Omega_{c}$} & \colhead{$L$ ($L_{\odot}$)}
}
\startdata
G9.62+0.19 & $29 \pm 3$ & $150 \pm 5$ & ($5.0\pm0.2$)$\times 10^{-6}$ & ($1.4\pm 0.2$)$\times10^{6}$ \\
G10.47+0.03 & $40\pm4$ & $150 \pm 5$ & ($5.0\pm0.2$)$\times 10^{-6}$ & ($2.1\pm0.4$)$\times10^{6}$ \\
G12.89+0.49 & $25\pm3$ & $150 \pm 5$ & ($1.0\pm0.2$)$\times 10^{-6}$ & ($3.7\pm0.5$)$\times10^{5}$ \\
G16.57-0.05 & $23\pm3$ & $160 \pm 5$ & ($4\pm1$)$\times 10^{-7}$ & ($2.6\pm0.4$)$\times10^{5}$ \\
G19.88-0.53 & $22\pm3$ & $160 \pm 5$ & ($3\pm1$)$\times 10^{-7}$ & ($1.5\pm0.3$)$\times10^{5}$ \\
\enddata
\end{deluxetable*}

We adopted a two-component modified blackbody model (or two-component radiative transfer model) for all of the five target sources, using archival data of {\it {Spitzer}}\footnote{\url{https://irsa.ipac.caltech.edu/frontpage/}} (3.6, 4.5, and 8\,$\micron$) and {\it {Herschel}}\footnote{\url{http://archives.esac.esa.int/hsa/whsa/}} (70, 160, 250, 350, and 500\,$\micron$).
Using these infrared archival data, we obtained the flux density values with 0.5 pc in diameter toward all of the sources; 19\farcs83, 9\farcs62, 34\farcs38, 21\farcs92, and 31\farcs25 for G9.62+0.19, G10.47+0.03, G12.89+0.49, G16.57-0.05, and G19.88-0.53, respectively.
The coordinates of the centeral positions of these analyses are summarized in Table \ref{tab:source}.
Table \ref{tab:fluxdensity} summarizes the obtained values of flux densities at each frequencies.
We estimated uncertainties of flux densities including assumed 20\% calibration errors and rms noise levels of archival images calculated in CASA.

We assume that emission of 250\,$\micron$ (1.199169832 THz) is optically thin, and used fluxes at this wavelength as a reference.
We calculated the optical depths of other wavelengths using the following formula:
\begin{equation}
\tau_{\nu} = \tau({250\,\micron}) \left(\frac{\nu}{1.199169832\, {\rm {THz}}}\right)^\beta.
\end{equation}
The spectral index ($\beta$) was treated as a free parameter with a range of 1.0-2.4 \citep{2018A&A...615A..18R}.
We searched for the best $\beta$ values for each source by changing this parameter in steps of 0.05; $\beta=1.5$ for all of the sources except for G19.88-0.53, where $\beta=1.75$ is applied.
Since the emission regions of the warm components should be smaller than those of the cold components, we multiplied the warm component by $\Omega_{w}/\Omega{c}$ ($\Omega$ is solid angle) to correct this effect. 
We searched for the best combination of three free parameters ($T_{c}$, $T_{w}$, $\Omega_{w}/\Omega_{c}$), changing with steps of 1 K, 5 K and $1\times10^{-7}$ for $T_{c}$, $T_{w}$, $\Omega_{w}/\Omega_{c}$, respectively.

Figure \ref{fig:sed} shows the modified blackbody models which reproduce the observed flux values best.
Table \ref{tab:sed} summarizes the results including total luminosities derived from the results of the best blackbody models, i.e., $L=4\pi d^2 \int S_{\nu} d\nu$.
The adopted source distances ($d$) are summarized in Table \ref{tab:source}.
The obtained values here are different from \citet{2018MNRAS.473.1059U}, because of different areas.
The source luminosities derived by our method are typical for MYSOs.

\section{Comparisons of model results with different cosmic ray ionization rates} \label{sec:a3}

\begin{figure*}[!th]
 \begin{center}
  \includegraphics[bb =0 20 945 894, scale=0.48]{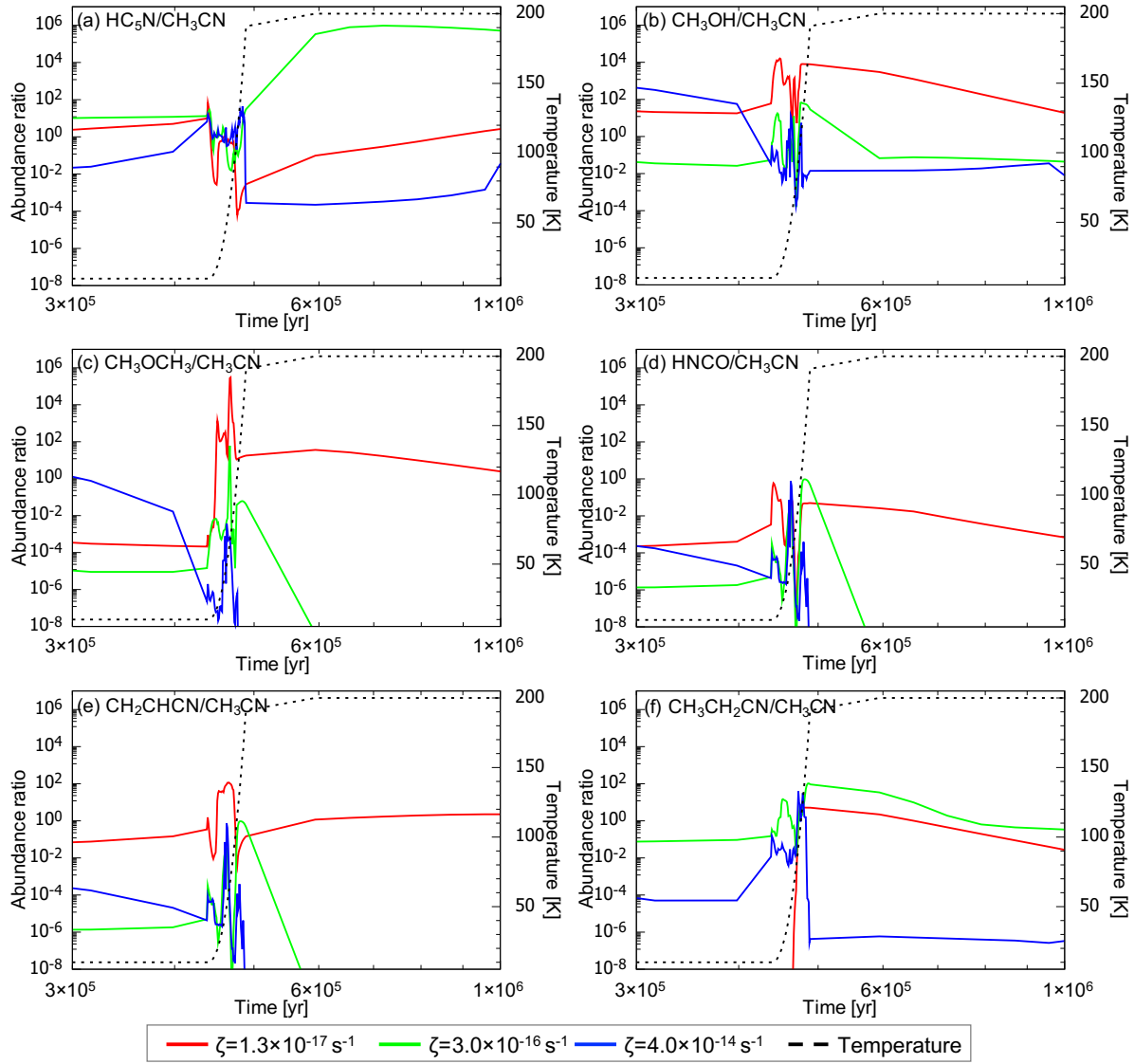}
 \end{center}
\caption{Comparisons of results obtained by chemical simulations with three different cosmic-ray ionization rates (red; $1.3 \times 10^{-17}$ s$^{-1}$, green; $3.0 \times 10^{-16}$ s$^{-1}$, blue; $4.0 \times 10^{-14}$ s$^{-1}$). The black dashed curves indicate the temperature. \label{fig:modelcomp}}
\end{figure*}

Figure \ref{fig:modelcomp} shows comparisons of abundances with respect to CH$_{3}$CN derived by chemical simulations with three different cosmic-ray ionization rates. 
Abundance ratios of O-bearing COMs (CH$_{3}$OH and CH$_{3}$OCH$_{3}$) and CH$_{2}$CHCN with respect to CH$_{3}$CN are lower in models with two higher cosmic-ray ionization rates.
The HC$_{5}$N/CH$_{3}$CN abundance ratio with $\zeta = 3.0 \times 10^{-16}$ s$^{-1}$ is much higher compared to the other two models.
Details of cyanopolyyne chemistry in different cosmic-ray ionization rates were explained in \citet{2019ApJ...881...57T}.
These models with the two higher cosmic-ray ionization rates cannot reproduce the observed abundance ratios simultaneously. 

{}



\end{document}